\documentclass[useAMS]{mn2e}
\usepackage{epsfig}
\usepackage{rotating}

\def\lesssim{\,\lower2truept\hbox{${<\atop\hbox{\raise4truept\hbox{$\sim$}}}$}\,}
\def\gtrsim{\,\lower2truept\hbox{${>\atop\hbox{\raise4truept\hbox{$\sim$}}}$}\,}

\title[The PACO bright sample]{The Planck-ATCA Co-eval Observations (PACO) project: the bright sample}
\author[M. Massardi et al.]{
\parbox[t]{\textwidth}
{Marcella Massardi$^{1}$\thanks{E-mail: marcella.massardi@oapd.inaf.it}, Anna Bonaldi$^{1,2}$, Laura Bonavera$^{3,4}$, Marcos L\'opez-Caniego$^{5}$, Gianfranco De Zotti$^{1,3}$,  Ronald D.\ Ekers$^4$}
\vspace*{8pt} \\
$^{1}$INAF, Osservatorio Astronomico di Padova, Vicolo dell'Osservatorio 5, I-35122 Padova, Italy\\
$^{2}$Jodrell Bank Centre for Astrophysics, School of Physics \& Astronomy, University of Manchester, Oxford Road, Manchester M13 9PL\\
$^{3}$SISSA, via Bonomea 265, I 34136 Trieste, Italy\\
$^{4}$Australia Telescope National Facility, CSIRO Astronomy and Space Science, PO Box 76, Epping, NSW 1710, Australia\\
$^{5}$Instituto de F\'isica de Cantabria (CSIC-UC), Avda. los Castros s/n, 39005 Santander, Spain}
\begin{document}

\date{}

\pagerange{\pageref{firstpage}--\pageref{lastpage}} \pubyear{2010}

\maketitle

\label{firstpage}

\begin{abstract}
The Planck-ATCA Co-eval Observations (PACO) have provided flux density measurements of well defined samples of AT20G radio sources at frequencies below and overlapping with Planck frequency bands, almost simultaneously with Planck observations. We have observed with the Australia Telescope Compact Array (ATCA) a total of 482 sources in the frequency range between 4.5 and 40 GHz in the period between July 2009 and August 2010. Several sources were observed more than once. In this paper we present the aims of the project, the selection criteria, and the observation and data reduction procedures. We also discuss the data in total intensity for a complete sample of 189 sources with $S_{\rm 20 GHz}>500$ mJy, Galactic latitude $|b|>5^\circ$, and declination $\delta<-30^\circ$, and some statistical analysis of the spectral behaviour and variability of this sample, referred to as the ``bright PACO sample''. Finally we discuss how these data could be used to transfer absolute calibrations to ground based telescopes using the CMB dipole calibrated flux densities measured by the Planck satellite, and we provide some test fluxes on bright calibrators.
\end{abstract}

\section{Introduction}
The ESA's Planck satellite\footnote{http://sci.esa.int/science-e/www/area/index.cfm?fareaid=17} is surveying the sky in nine frequency bands (30, 44, 70 GHz for the Low Frequency Instrument, LFI, and 100, 143, 217, 353, 545, 857 GHz for the High Frequency Instrument, HFI, with the beam FWHM ranging from 33 to 5 arcmin, Planck Collaboration 2011c). At the HFI frequencies it will provide the first all-sky surveys ever, while at LFI frequencies its higher sensitivity and resolution will allow a significant improvement over WMAP (Gold et al. 2011, Planck Collaboration 2006 and 2011c, Leach et al. 2008, Massardi et al. 2009). Planck thus offers a unique opportunity to carry out an unbiased investigation of the spectral properties of radio sources in a poorly explored frequency range, partially unaccessible from the ground.

Planck's science yields can be greatly increased by simultaneous (i.e. not affected by variability) ground based observations at lower frequencies as well as at frequencies overlapping with Planck channels. Simultaneous observations at overlapping frequencies are important to transfer the Planck flux density calibration, which is based on the CMB dipole, to the primary calibrators used by the ground-based radio telescopes. These observations can also be used to help with the validation of sources detected by Planck.

At present, the flux density scales for ground-based telescopes are well determined for frequencies up to 10 GHz but there are known problems at the higher frequencies. The difficulty occurs because the radio interferometers, which can make the most precise flux density measurements at high frequencies, need point sources to calibrate and the strong point sources are mostly highly variable AGNs. This has led to the use of the planets, Mars and Uranus, as primary calibrators. These planets are resolved at all but the shortest interferometer baselines at the higher frequencies and their flux densities are estimated using relatively complex atmospheric models. In the case of Uranus there is a further complication because of the known orbital change in flux density as our view changes from equatorial to polar during the Uranian year. Simultaneous observations of very bright sources give a unique opportunity to tie all the high frequency flux density scales together with the Planck's calibration. It should be noted however that the source flux densities given in the Early Release Compact Source Catalogue (ERCSC, Planck collaboration 2011b) are averages over all observations, that may extend over several months. Therefore they may be used for some statistical purposes, as in Planck Collaboration (2011d) and in papers in preparations from our group, while for calibration purposes we need to use the Planck time ordered data, averaged over shorter periods, comparable with the timescales of ground based observations.

Source confusion is a serious issue for Planck, due to its rather large beam sizes. Sources near Planck's detection limit may be strongly affected by the Eddington's (1913) bias that leads to systematic flux density overestimates (Hogg \& Turner 1998). Also, confusion fluctuations may produce high intensity peaks that may be misinterpreted as real sources. Moreover, confusion can shift the positions of intensity peaks from the true source positions (Hogg 2001), complicating their  identification. Simultaneous ground based observations, with much better resolution and signal-to-noise ratio will allow an accurate control of these effects.

Source Spectral Energy Distributions (SEDs) over a frequency range as large as possible are crucial in determining their physical properties and in identifying the different components that may contribute to their emission. Given that observations in the full Planck frequency range will not be repeated at least in the foreseeable future, it is essential not to \textbf{lose} the opportunity for near simultaneous observations while Planck is flying.

Observations simultaneous with Planck can be planned only for previously known sources. The Australia Telescope 20 GHz (AT20G; Murphy et al. 2010; Massardi et al. 2010) survey provides the best sample for this purpose. This is the largest ground-based sample of the high radio frequency ($>10$ GHz) sky: it is 93\% complete above 100 mJy at 20 GHz over the whole Southern sky with follow-up (within a few weeks time) at 4.8 and 8.6 GHz. For typical source spectra (Massardi et al. 2008), it is somewhat deeper than Planck, that is expected to reach detection limits ranging from $\simeq 400\,$mJy at 30 GHz to $\simeq 220\,$mJy at 100 GHz (Leach et al. 2008).

The above considerations have motivated the Planck-ATCA Co-eval Observations (PACO) project, described in \S\,\ref{sec:PACO}. The observations exploited the capabilities of the new Australia Telescope Compact Array Broadband Backend (CABB; Ferris \& Wilson 2002, Koribalski et al. in prep.) system between 4.5 and 40 GHz.
\section{The PACO project}\label{sec:PACO}
\begin{figure}
    \hspace{-2cm}
  \includegraphics[width=10cm, height=12cm, angle=90]{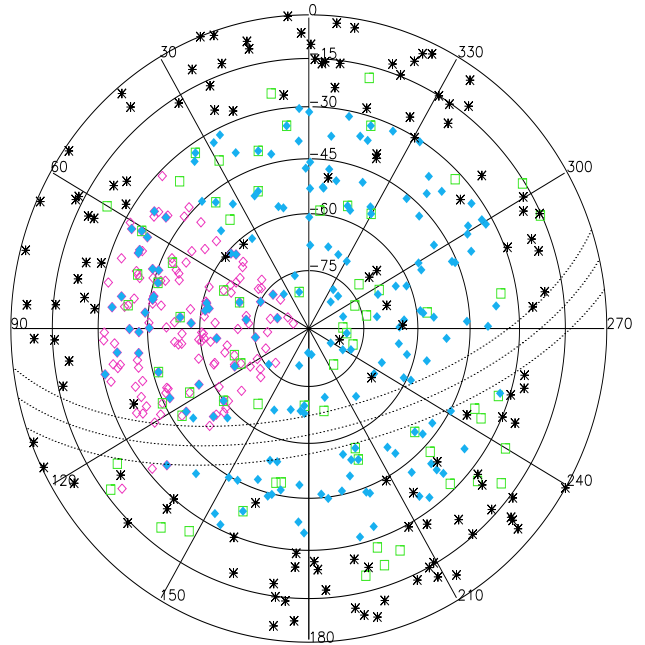}
  \caption{Polar projection of the Southern sky, showing the distribution of the PACO samples (RA=0h is at the top and the RA increases counterclockwise). The dotted lines indicate the Galactic plane and bound the area with Galactic latitude $|b|<5^\circ$. Sources in these regions were excluded from the PACO samples. The faint PACO sample (open pink diamonds) occupies the area with $3\hbox{h}<{\rm RA}<9$h. The bright PACO sample (filled blue diamons) covers the whole area at $\delta<-30^\circ$. The spectrally-selected sample (green squares), the blazars and the AT calibrator sources (black asterisks) are spread over the whole southern sky.} \label{fig:map}
\end{figure}
\subsection{Sample selection} \label{sec:sample}

The PACO sample is a combination of several sub-samples selected in order to maximize the scientific outcome in the allocated time. The ATCA allows us to make
observations at frequencies lower than or close to those of the lowest frequency
Planck channels at 30 and 44 GHz.
We have observed 482 AT20G extragalactic sources at Galactic latitude
$|b|>5^\circ$ and outside a $5^\circ$ radius circle centered at
$\hbox{RA}=80.8939^\circ$, $\delta=-69.7561^\circ$ (the LMC region).
Of these, 344 sources form 3 partially overlapping complete sub-samples, selected for different purposes.
\begin{itemize}
\item The ``faint PACO sample'' is made of the 159 sources with $S_{20{\rm GHz}}>200$ mJy in the southern ecliptic pole region (ecliptic latitude $<-75^\circ$) and with $3\hbox{h}<\hbox{RA}<9\hbox{h}$, $\delta<-30^\circ$ (Bonavera et al. 2011). Near the ecliptic poles Planck's scan circles intersect. Therefore the area is covered many times, and Planck's sensitivity is maximal. Nevertheless, we expect that many sources in this sample will not be detected by Planck. On the other hand, Poisson fluctuations due to extragalactic sources below the detection limit (expected to be of 100--250 mJy in the range 100--353 GHz, Leach et al. 2008) are the main contaminant of Planck maps on angular scales smaller than $20'$--$30'$ at frequencies of up to $\simeq 100\,$GHz (Toffolatti et al. 1998, 1999; De Zotti et al. 1999). It is therefore important to characterize as accurately as possible radio sources down to below the Planck detection limits at the time of Planck observations.

\item The ``spectrally-selected PACO sample'' comprises the 69 sources with $S_{20\rm{GHz}}>200$ mJy and spectra classified by Massardi et al. (2010) as inverted or upturning in the frequency range 5--20 GHz, selected over the whole southern sky.

\item The ``bright PACO sample'' comprising the 189 sources with $S_{20\rm{GHz}}>500\,$mJy at $\delta<-30^\circ$, is presented and discussed in this paper.

\item An additional sample (partially overlapping with those defined above) is made of 63 bright southern blazars included in a long-term (up to 15 years) monitoring program at various frequencies with the Swedish-ESO Submillimetre Telescope (SEST, Tornikoski et al. 1996) and observed also with the LABOCA instrument on the APEX. Based on a source counts analysis, this sample appears to be 91 per cent complete down to the cutoff of 1 Jy at 90 GHz. These sources are of special interest because some of them may be flaring during the Planck mission. If this happens, we get an unusually broad simultaneous frequency coverage of the flare.
\end{itemize}
The samples described above include 76 ATCA calibrators with 20 GHz flux density $S_{20\rm{GHz}}>200$ mJy that showed more than 10\% variability at this frequency in the last three years. Further 121 variable ATCA calibrators were added to the PACO sample, so that their total observed number is of 197. Our data will also be included in the AT calibrator database.

Our observations allow us to characterize the source spectra in the range 4.5--40 GHz. Planck data will allow us to extend the spectral coverage up to 857 GHz, at least for the brightest flat-spectrum sources.

The classification as extragalactic objects and an indication of source extendedness up to 20 GHz were taken from the AT20G catalogue. AT20G positions were also used unless more precise coordinates were available, as is the case, e.g., for the AT calibrators whose positions are based on VLBI observations. A map of our sample is shown in Fig. \ref{fig:map}. As pointed out by Massardi et al. (2010) the sample may still contain some Galactic sources, but they should be very few because the relatively high resolution 20 GHz observations preferentially select compact objects ($<30$ arcsec) and Galactic sources with $|b|>1.5^\circ$ are rarely that small, especially at the brighter flux density levels.

\subsection{Observations} \label{sec:observations}

\begin{table}
\caption{The flux density values of the model for the primary calibrator PKS~B1934-638 at the center of our observing frequency bands, as it is in the Miriad version we used.}\label{tab:1934}
\begin{tabular}{cc}
\hline
Frequency   &  Flux density\\
$[$GHz$]$   &  $[$mJy$]$   \\
\hline
    5.5 & 4965\\
    9.0 & 2701\\
   18.0 & 1101\\
   24.0 &  725\\
   33.0 &  441\\
   39.0 &  335\\
\hline
\end{tabular}
\end{table}
\begin{figure}
  \includegraphics[width=5.5cm, angle=90]{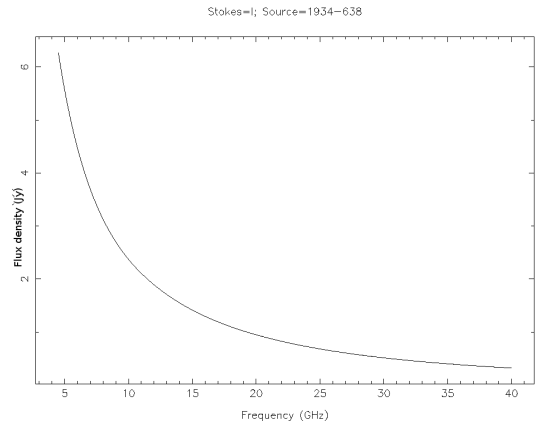}
  \caption{Model for PKS~1934-638 in the frequency range between 4.5 and 40 GHz, as it is in the Miriad version we used.} \label{fig:1934}
\end{figure}
\begin{table*}
\caption{The 63 good weather observational runs for the PACO project between July 2009 and August 2010.}\label{tab:summary_obs}
\begin{tabular}{lccllccl}
\hline
Date &  array&  Allocated & Frequency &  Date &  array&  Allocated & Frequency \\
     &       &  time [h]  & Range       &       &       &  time [h]  & Range       \\ \hline
2009-07-09 & H75C&  8.5 & 4.5-40 GHz    &     2010-02-07 &  6A & 4.5  & 4.5-25 GHz \\
2009-07-10 & H75C&  5   &  17-40 GHz    &     2010-02-14 & 6A  & 9    & 4.5-25 GHz \\
2009-07-16 & H75C&  2   & 4.5-25 GHz    &     2010-02-28 &750B & 8.5  & 4.5-40 GHz \\
2009-07-18 & H75C&  1   &  32-40 GHz    &     2010-03-02 &750B & 4    & 4.5-25 GHz \\
2009-07-25 & H75C&  8   & 4.5-40 GHz    &     2010-03-06 &H168 & 10   & 4.5-40 GHz \\
2009-07-26 & H75C&  7   & 4.5-40 GHz    &     2010-03-15 &H168 &  8   & 4.5-40 GHz \\
2009-07-31 & 1.5A&  7   & 4.5-40 GHz    &     2010-04-01 & 6A  &  12  & 4.5-40 GHz \\
2009-08-09 &  6D &  6.5 & 4.5-40 GHz    &     2010-04-06 & 6A  &  4  &  4.5-25 GHz \\
2009-08-13 &  6D &  8   & 4.5-40 GHz    &     2010-04-10 & 6A  &  5   & 4.5-40 GHz \\
2009-08-24 &  6D &  2.5 &  17-40 GHz    &     2010-04-11 & 6A  &  6.5 & 4.5-40 GHz \\
2009-08-25 &  6D &  3.5 &  17-40 GHz    &     2010-04-23 & 6A  &  7   & 4.5-40 GHz \\
2009-08-30 &  6D &  6   & 4.5-40 GHz    &     2010-04-24 & 6A  &  7   & 4.5-40 GHz \\
2009-09-15 & H214&  8   & 4.5-40 GHz    &     2010-05-01 & 6A  &  4   & 4.5-40 GHz \\
2009-09-26 &  H75&  14  & 4.5-40 GHz    &     2010-05-03 & 6A  &  4   & 4.5-40 GHz \\
2009-09-29 &  H75&  4   & 4.5-40 GHz    &     2010-05-11 & H214&  4   & 4.5-40 GHz \\
2009-10-07 &  H75&  6.5 & 4.5-40 GHz    &     2010-05-18 & 6C  &  7   & 4.5-40 GHz \\
2009-10-14 & H168& 10   & 4.5-40 GHz    &     2010-05-30 & 6C  &  12  & 4.5-40 GHz \\
2009-10-28 & 1.5B&  4   & 4.5-40 GHz    &     2010-06-05 & 6C  &  7.5 & 4.5-40 GHz \\
2009-10-29 & 1.5B&  4   & 4.5-40 GHz    &     2010-06-11 & 6C  &  4   & 4.5-40 GHz \\
2009-11-06 & 1.5B& 14.5 & 4.5-40 GHz    &     2010-06-14 & 6C  & 12   & 4.5-40 GHz \\
2009-11-11 & 6B  &  4   & 4.5-40 GHz    &     2010-06-19 & 6C  & 14.5 & 4.5-40 GHz \\
2009-11-19 & 6B  & 11   & 4.5-40 GHz    &     2010-07-02 & 6C  &  4   & 4.5-40 GHz \\
2009-11-24 & 6B  &  4   & 4.5-40 GHz    &     2010-07-12 & 1.5D&  5.5 & 4.5-40 GHz \\
2009-12-04 &EW352&  14  & 4.5-40 GHz    &     2010-07-16 &EW352&  5   & 4.5-40 GHz \\
2009-12-07 &EW352&  13  & 4.5-40 GHz    &     2010-07-31 & H168&  4   & 4.5-40 GHz \\
2009-12-17 & 6A  &  9.5 & 4.5-40 GHz    &     2010-08-10 & H168&  6   & 4.5-40 GHz \\
2009-12-29 & 6A  & 8    & 4.5-40 GHz    &     2010-08-13 & H168&  2   & 32-40 GHz  \\
2010-01-03 & 6A  & 4.5  & 4.5-25 GHz    &     2010-08-14 & H168&  6.5 & 4.5-40 GHz \\
2010-01-17 & 6A  & 4    & 4.5-25 GHz    &     2010-08-15 & H168&  6.5 & 4.5-40 GHz \\
2010-01-24 & 6A  & 7    & 4.5-40 GHz    &     2010-08-20 & H168&  10  & 4.5-40 GHz \\
2010-01-27 & 6A  & 4    & 4.5-25 GHz    &     2010-08-21 & H168&  10  & 4.5-40 GHz \\
2010-02-06 & 6A  & 12   & 4.5-25 GHz    &                &     &      &            \\

\hline
\end{tabular}
\end{table*}
The Planck scanning strategy has been publicly released soon after the satellite launch. The satellite scans the sky in circles passing close to the ecliptic poles. All Planck receivers cross the position of sources lying on the ecliptic equator within a few days. Sources close to the ecliptic poles remain within the satellite focal plane for up to 2.5 months. The satellite covers the whole sky in about 6 months. Because the maps are the result of averaging all the scans through a position, we can't get information from them on the very short term variability. Variability tests on the time ordered data are limited to the brightest objects because of lack of integration on source. We consider observations to be `co-eval' with the satellite if performed within 10 days from the satellite observations at any of its LFI frequencies. The Planck-On-the Flight forecaster (Massardi \& Burigana 2010) has been developed and used to predict when the PACO sources are observed by the satellite, according to its pre-programmed pointing lists and the focal plane properties.

The new ATCA CABB system allows $2\times2$ GHz simultaneous bands in continuum. Applied to the $6\times 22\,$m ATCA antennas it gives a noise level of 0.5\,mJy in 1\,min on source in the 7\,mm band. We choose to use the 7\,mm receivers with bands centered at 33 and 39\,GHz, to overlap the lower-frequency Planck channels, the 12\,mm receiver with bands centered at 18 and 24\,GHz, to include the AT20G selection frequency, and we extended the SEDs to lower frequencies using the 3-6\,cm receiver with bands at 5.5 and 9\,GHz.

The AT20G catalogue is dominated by compact flat spectrum sources so the AT20G selection guarantees that most radio sources have small angular sizes so that the particular configuration of the ATCA is not crucial for flux density estimation.
Hybrid arrays are preferable for equatorial sources, for which the beam shape of EW arrays is elongated. Because of the risk of confusion with the highly elongated synthesized beam for sources with declination $\delta>-10^\circ$, they have not been observed at 3-6\,cm. This is not a concern at shorter wavelengths where there is no confusion in the smaller primary beam.

The scheduling process tries to guarantee the simultaneity with the Planck observations following the satellite scanning strategy and to minimize the slewing time, maximize the time on source and complete the measurements at the selected frequencies (i.e. including considerations about the weather, the array configuration in use and the minimization of the band switches).

For each epoch and each frequency, a very bright compact object is used for bandpass calibration. Observations of the bandpass are usually repeated several times to account for time instability of the band. PKS~B1934-638 is the primary calibrator for all the frequencies. The spectral model for the primary calibrator that we have used in all the epochs is in Fig.~\ref{fig:1934}. Tab.~\ref{tab:1934} lists the model flux density values at the center of our observing frequency bands.

At 7 mm we have sometimes observed Uranus. Although the planet is always resolved at our longer baselines, we can use the shorter baselines to transfer the calibration scale. By comparing with the PKS~1934-638 flux densities, we noted that the model for Uranus requires a flux density correction by a factor of $\sim0.82$ at 33 GHz and $\sim0.84$ at 39 GHz. This is in agreement with what estimated on the basis of the on-going calibrator model update, as the absolute flux density scale at this frequency is still under investigation at both ATCA and VLA (J. Stevens, priv. comm., see \S \ref{sec:abscalib}). Hence we applied these correction factors to all the 7\,mm flux densities calibrated with the planet.

A leakage calibrator selected among the polarized objects of the AT20G (a bright point-like unpolarized one should work as well) is observed at least 3 times, better if across the transit of the source, at each frequency for almost all the observing runs.
Pointing calibration is performed regularly for 12 and 7\,mm observations. Each source at each frequency is observed for 1.5\,min in a single pointing.

Thanks to the compactness and high signal to noise of most of the PACO sources neither imaging nor phase calibration is necessary. Since sources are self-calibrated we get good images when observing with hybrid arrays. We got a suitable flux density estimation from the visibilities using the triple-correlation techniques, as will be explained in the next sections.

The project got $\sim$450 hours allocated over 65 epochs between July 2009 and August 2010 (i.e. during the time when Planck completed two all-sky surveys), of which 2 epochs have been discarded for bad weather. Because of the high sensitivity in total intensity and the brightness of the sources in our sample we could observe on average 20 objects per run at all the 3 pairs of frequencies. Bad weather seriously affected some of the Austral summer epochs making it impossible to get the 7 mm data in a few runs. Table~\ref{tab:summary_obs} summarizes the configurations, and frequencies observed in each run.

\subsection{Data reduction} \label{sec:reduction}
\begin{figure}
\begin{center}
\hspace{-1.5cm}
\includegraphics[scale=0.4]{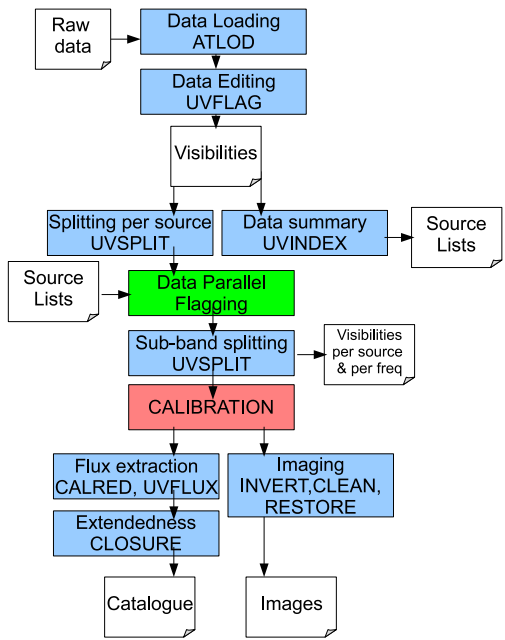}
\caption{Data reduction pipeline diagram. In capital letters are the Miriad tasks used.}\label{fig:datareduction}
\end{center}
\end{figure}

\begin{figure}
\hspace{-1.5cm}
\includegraphics[scale=0.4]{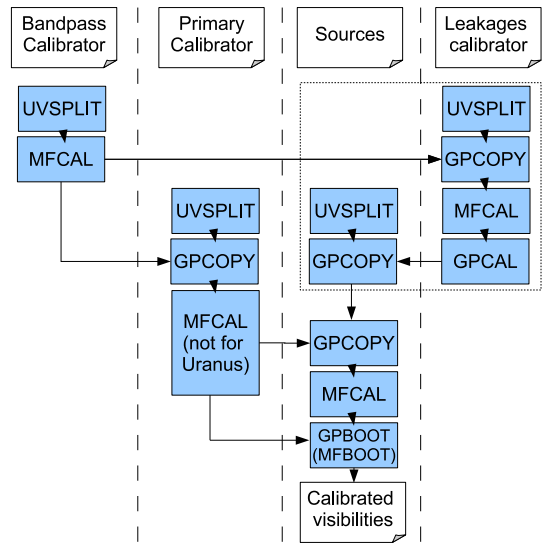}
\caption{Scheme of the calibration performed with MIRIAD tasks. The indications in parenthesis refer to the case of Uranus as primary calibrator. The steps included in the dotted square have been applied only when a suitable leakage calibrator has been observed.}\label{fig:calibration}
\end{figure}

Data reduction is performed by a C-shell coded pipeline which uses tasks\footnote{A comprehensive description of each MIRIAD task can be found at the website http://www.atnf.csiro.au/computing/software/miriad/taskindex.html} from the MIRIAD software package (Sault et al. 1995) according to the scheme in Fig.~\ref{fig:datareduction}. Chunks of data affected by bad weather conditions or instrumental malfunction are removed, as well as data affected by shadowing or known radio interference. Automatic procedures identify and flag among the 2048 channels, which constitute the 2.048 GHz bandwidth for each IF, those that exceed by 5$\sigma$ the mean value for each baseline for each source. Typically less than 1\% of the band is flagged. In order to properly define the detailed source spectral behaviour we have split each 2 GHz band into $4\times 512$ MHz sub-bands, and calibrated each sub-band independently. The steps for calibration are plotted in Fig.~\ref{fig:calibration}.

\subsubsection{Extended sources}

We expect that the AT20G analysis of extended sources have already identified all the extended sources up to 20GHz. We have therefore done the analysis only for the 7\,mm data, applying the automatic procedure to distinguish between point-like and extended sources used by Massardi et al. (2008) for the AT20G bright sample.

The ratio between the actual and the theoretical phase closure rms calculated accounting for the noise in the beam gives an indication on whether the source is  extended. For a point source the phase closure rms should be zero and is independent of instrumental effects. When the ratio of actual to theoretical phase closure rms is larger than 3 the source should be considered extended. We have used this ratio as an indication of extendedness for the sources that appear to be extended at 7\,mm and were not identified as extended in the AT20G catalogue.

\subsubsection{Flux density estimation}\label{sec:flux}

To estimate the flux densities for point sources we have used the triple product (i.e. the geometrical average of the amplitude over a closure triangle, averaged over all the possible triangles, given by the MIRIAD task CALRED). Our flux density estimation techniques are well suited for point sources. We adopted the amplitude of the shortest spacings for the array configuration used (see Tables~\ref{tab:summary_obs} and \ref{tab:config_size}), given by the MIRIAD task UVFLUX, as the best flux estimator. This approach still underestimates the flux densities for sources more extended than $15(100\hbox{m}/b_{\rm min})(39\hbox{GHz}/\nu)$ arcsec, where $b_{\rm min}$ is the shortest baseline for any used configuration. Thus our flux densities are increasingly underestimated and, as a consequence, our samples are increasingly incomplete for more and more extended sources. Table \ref{tab:config_size} lists the shortest baselines ($b_{\rm min}$), and the primary and synthesized beam sizes for the central frequencies of our observing runs for all the array configurations used.

\begin{table*}
\caption{The array configurations and their resolution properties.}\label{tab:config_size}
\begin{tabular}{lcccccc}
\hline
Array         & Shortest     & frequency    &Primary beam & Beam IF1              & Beam IF2                & $D_{\rm max}$ \\
Configuration & Baseline [m] & IF1 - IF2[GHz] &FWHM [arcmin]& [arcsec]             & [arcsec]               & [arcmin] \\
\hline
H75c          & 31 (1-4,2-4) & 5.5 - 9      & 8.7 - 5.3   & 115.51 x 115.51      &   70.59 x 70.59      & 6.0 - 3.7\\
              &              & 18  - 24     & 2.6 - 2.0   & 35.29 x 35.29      &   26.47 x 26.47      & 1.8 - 1.4\\
              &              & 33  - 39     & 1.5 - 1.2   & 19.25 x 19.25      &   16.29 x 16.29      & 1.0 - 0.9\\
H168d         & 61 (1-4,2-4) & 5.5 - 9      & 8.7 - 5.3   & 50.23 x 50.23      &   30.69 x 30.69      & 3.1 - 1.9\\
              &              & 18  - 24     & 2.6 - 2.0   & 15.35 x 15.35      &     11.51 x 11.51    & 0.9 - 0.7\\
              &              & 33  - 39     & 1.5 - 1.2   &  8.37 x 8.37       &  	 7.08 x 7.08     & 0.5 - 0.4\\
H214c         & 82 (2-4)     & 5.5 - 9      & 8.7 - 5.3   & 38.9 x 38.9        &    23.78 x 23.78     & 2.3 - 1.4\\
              &              & 18  - 24     & 2.6 - 2.0   &  11.89 x 11.89     &    8.92 x 8.92       & 0.7 - 0.5\\
              &              & 33  - 39     & 1.5 - 1.2   &  6.48 x 6.48      &     5.49 x 5.49      & 0.4 - 0.3\\
1.5A          & 153 (1-2)    & 5.5 - 9      & 8.7 - 5.3   &  7.2 x 8.31        &   	 4.4 x 5.08     & 1.2 - 0.7\\
              &              & 18  - 24     & 2.6 - 2.0   &  2.2 x 2.54        &   1.65 x 1.91        & 0.4 - 0.3\\
              &              & 33  - 39     & 1.5 - 1.2   &  1.2 x 1.39        &   1.02 x 1.17        & 0.2 - 0.2\\
1.5B          &  31 (1-2)    & 5.5 - 9      & 8.7 - 5.3   &  7.2 x 8.31        &   4.4 x 5.08        & 6.0 - 3.7\\
              &              & 18  - 24     & 2.6 - 2.0   &  2.2 x 2.54        &   1.65 x 1.91        &1.8 - 1.4\\
              &              & 33  - 39     & 1.5 - 1.2   &  1.2 x 1.39        &   1.02 x 1.17        &1.0 - 0.9\\
6B            & 214 (4-5)    & 5.5 - 9      & 8.7 - 5.3   &  2.48 x 2.86       &   1.51 x 1.75        & 0.9 - 0.5\\
              &              & 18  - 24     & 2.6 - 2.0   &  0.76 x 0.87      &    0.57 x 0.65       &   0.3 - 0.2\\
              &              & 33  - 39     & 1.5 - 1.2   &  0.41 x 0.48      &    0.35 x 0.4        &   0.1 - 0.1\\
6D            & 77 (4-5)     & 5.5 - 9      & 8.7 - 5.3   &  2.48 x 2.86       &   1.51 x 1.75        & 2.4 - 1.5\\
              &              & 18  - 24     & 2.6 - 2.0   &  0.76 x 0.87       &   0.57 x 0.65        & 0.7 - 0.6\\
              &              & 33  - 39     & 1.5 - 1.2   &  0.41 x 0.48       &   0.35 x 0.4         & 0.4 - 0.3\\
\hline
\end{tabular}
\end{table*}

The noise term of the flux density errors for point sources are calculated as the rms of the triple product amplitudes over all the possible triangles of antennas. For extended sources the rms of the triple product has been multiplied by $\sqrt{n_b}$ where $n_b$ is the number of baselines of the used configuration. An analysis of the gain term of the flux density error including systematics and calibration uncertainties is presented in \S\,\ref{sec:errors}.

\subsection{Quality assessment of the results}
By running the data reduction pipeline previously described we have generated a preliminary catalogue for each observing epoch. This catalogue was then processed through a quality assessment procedure. Depending on the results of the assessment, we either proceeded to create the final catalogue for the considered epoch or we carried out further investigations in order to improve the results.

The quality assessment step is needed because each observation can be affected by different problems (bad weather conditions, some antenna correlations to be rejected, residual spikes, etc.) which could not be addressed by the automatic pipeline. It is designed to identify the major flaws the catalogue of each observing epoch may have and to perform an a-posteriori flagging of the bad data.

The quality check assumes that the true SEDs do not contain discontinuities; those  observed are interpreted as due to residual spikes in the data or to calibration errors. The smoothness of the spectra is checked by comparing the data points with a fitting polynomial function with degree equal to the number of 2 GHz bands observed. We stress that, at this stage, the fit is done for flagging purposes only.

For each SED of the considered epoch we compute the fractional rms divergence, $\sigma$, of the data points from the fitting line. Next we exclude those points (if any) that are more than 4$\sigma$ off and recompute the fractional divergence over all the other points. The procedure is iterated until the value of $\sigma$ converges. To get rid of particularly bad epochs, if the final $\sigma$ is higher than 0.02, which corresponds to $\simeq 2 \times$ the median rms fractional divergence for all the observing epochs, we do one more iteration with $\sigma=0.02$.
Whenever two data points within the same 2 GHz band are discarded, it is likely that some problems are affecting the whole band, and we flag it for the considered source.

When the fraction of good data for a given observing epoch exceeds 90\% (and this happens in most cases) we directly create the final catalogue of the unflagged data. All the other cases are inspected in order to identify the problems and new runs of the data reduction pipeline with improved specifications are performed aiming at reducing the data loss to no more than 10-15\%. The recovery succeeded in most cases. The exceptions are due to problems affecting observations of the flux density calibrator for some bands. In a few cases the calibrator couldn't be recovered to a satisfactory level and the whole target sample was rejected. After the quality assessment process, we are left with $\sim85$\% of our data.

\subsubsection{Flux density error estimation}\label{sec:errors}

The gain term of the noise for each epoch was estimated as the rms fractional divergence between the data and the fit, re-calculated as discussed in the previous section after all the flagged data have been removed. For the days with less than 10 good fits available we have used $\sigma_g=0.012$, equal to the median of the rms fractional divergence calculated over all the epochs for point sources only. The gain term, as we calculated it, includes also flux density errors due to pointing errors. Hence, the final error bars for each point-source in each epoch have been estimated as square root of the square of the gain term multiplied by the source flux density plus the square of noise term (estimated as discussed in \ref{sec:flux}). Since our flux density estimation techniques are not well suited for extended sources, we have increased their gain term used in the error estimate to a minimum of $\sim5$ per cent of their flux density. This value has been obtained by scaling the rms fractional divergence calculated over all the epochs by $\sqrt{r_b}$ where $r_b=15/1$ is the ratio between the number of baselines used in case of point sources and in case of extended sources. This term does not account for any missing flux density of extended objects.

The gain term does not include any scaling error due to deficiencies in the absolute calibration errors which should affect equally all the frequencies calibrated with the same calibrator source (generally PKS~1934-638). The comparison with Planck CMB-dipole calibrated flux densities will allow us to estimate this bias and correct it.

\begin{figure*}
  \includegraphics[width=11cm, angle=90]{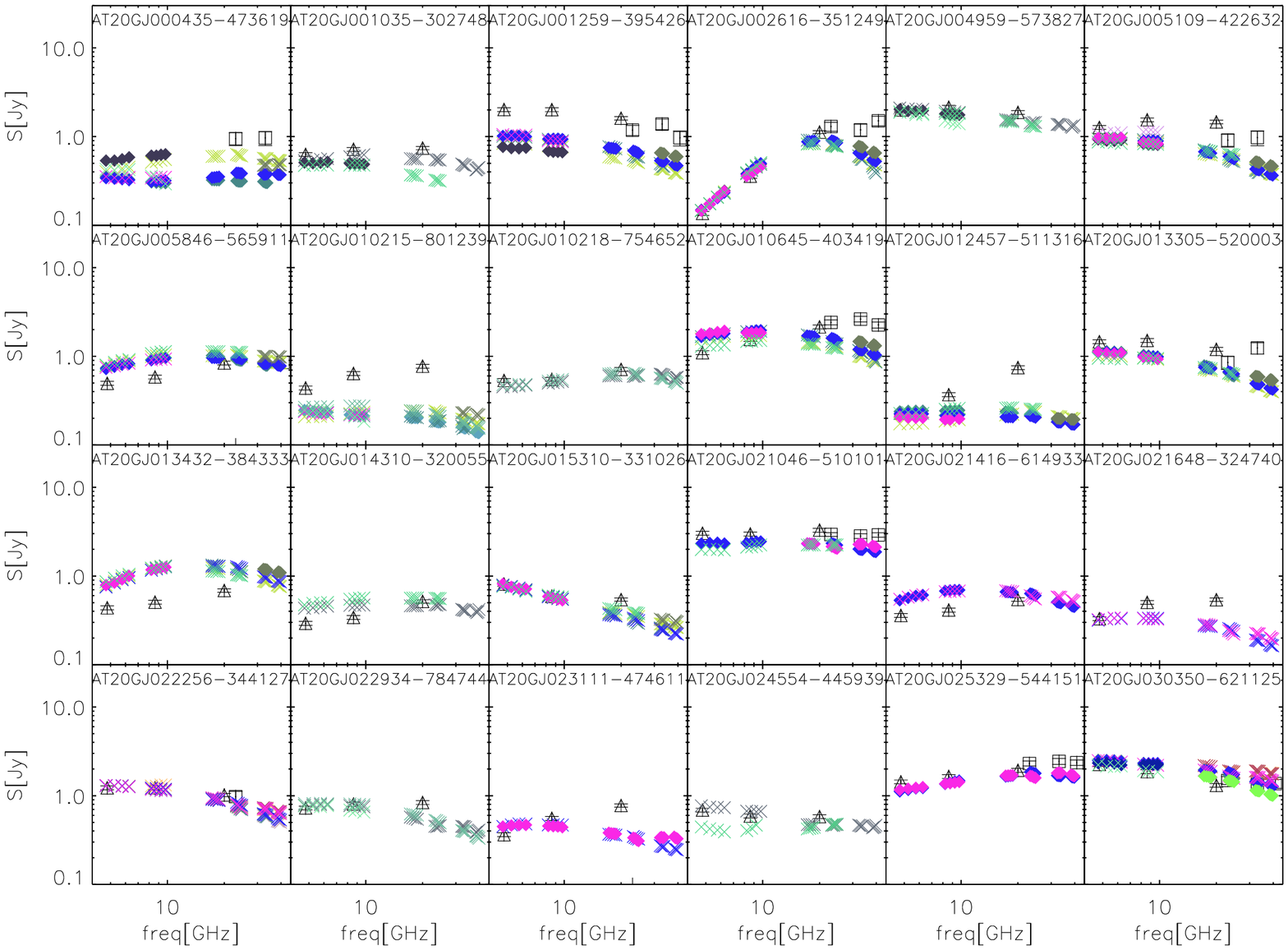}
  \includegraphics[width=11cm, angle=90]{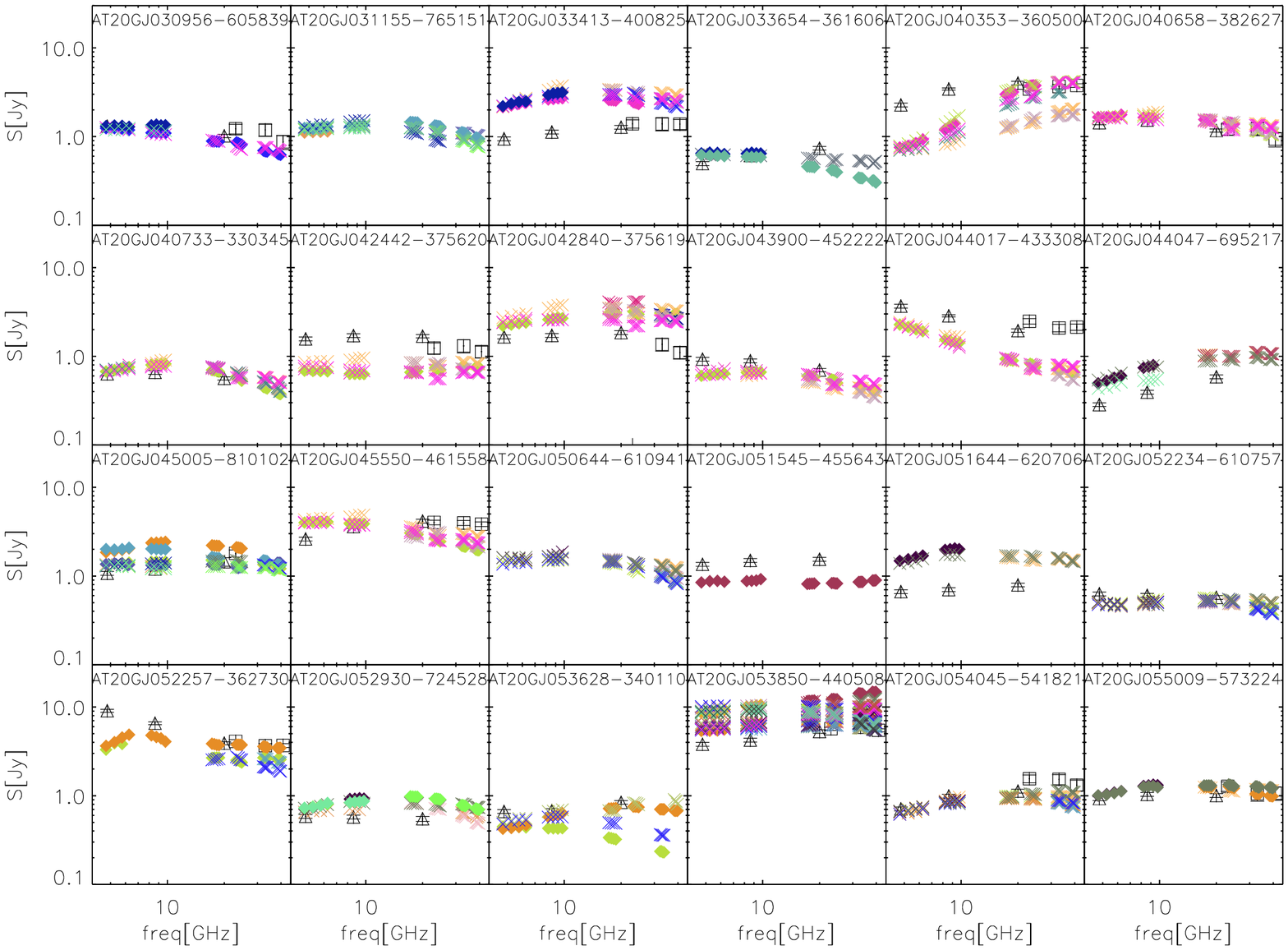}
 \caption{Spectra of sources in the PACO bright sample (crosses). Different colours correspond to different observing epochs. Diamonds corresponds to epochs simultaneous to Planck observations. To avoid overcrowding we have not plotted the error bars that are in all cases smaller than the symbol size. Also shown, for comparison, are the AT20G data (triangles) and, when available, WMAP data taken from the NEWPS\_5yr catalogue (squares; Massardi et al. 2009).} \label{fig:SED}
\end{figure*}
\addtocounter{figure}{-1}
\begin{figure*}
  \includegraphics[width=11.5cm, angle=90]{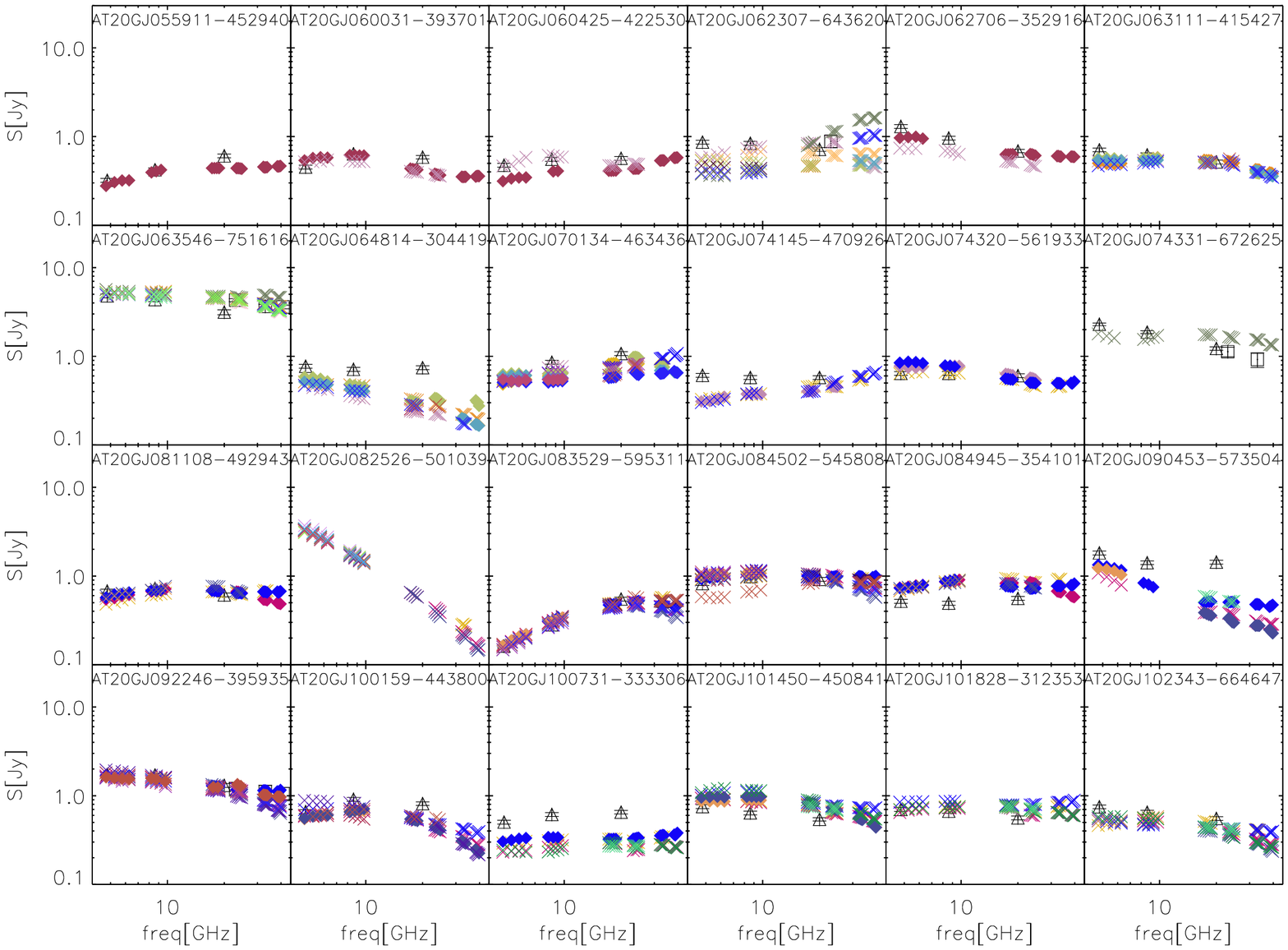}
  \includegraphics[width=11.5cm, angle=90]{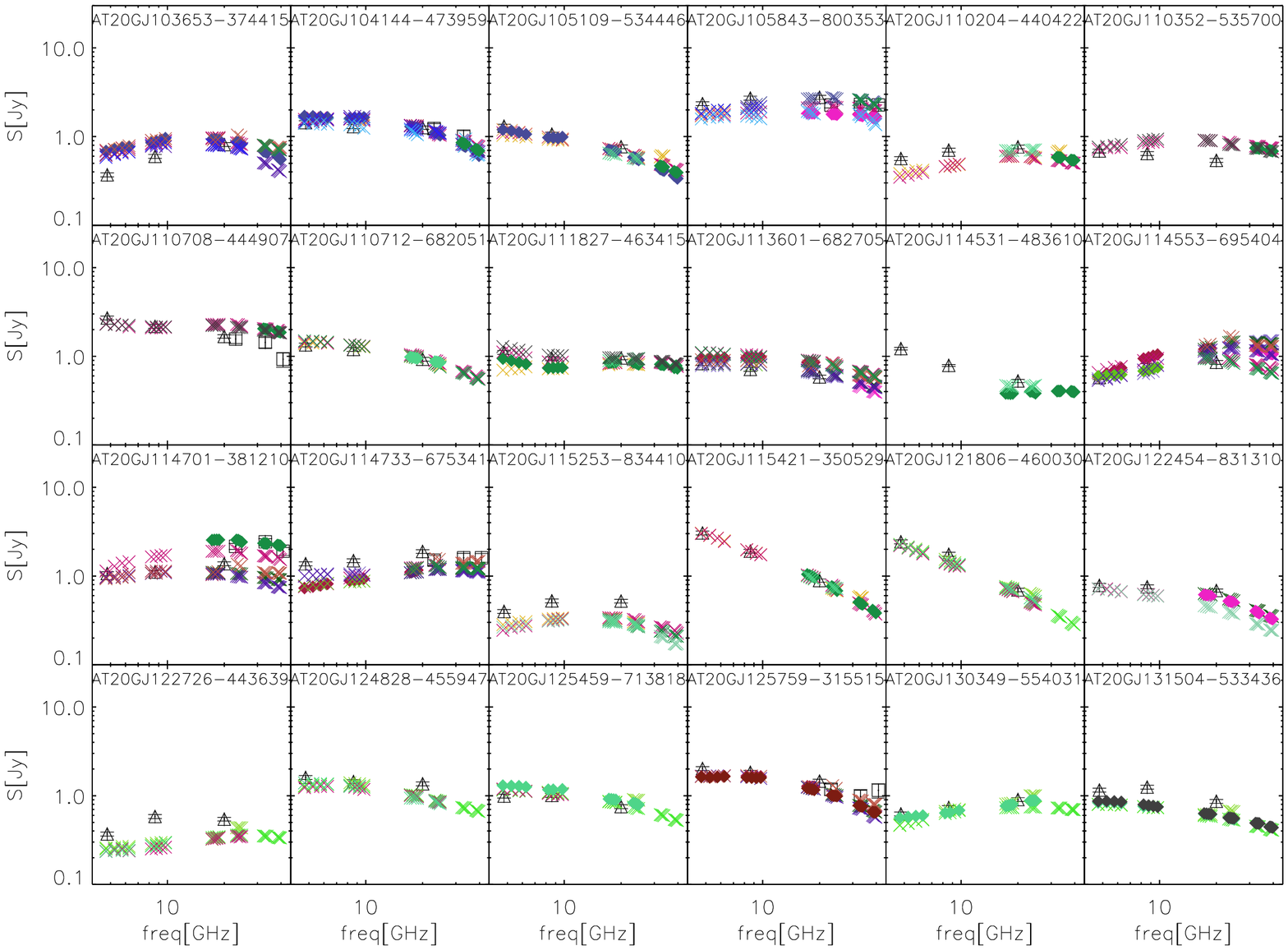}
  \caption{Continued.}
\end{figure*}
\addtocounter{figure}{-1}
\begin{figure*}
  \includegraphics[width=11.5cm, angle=90]{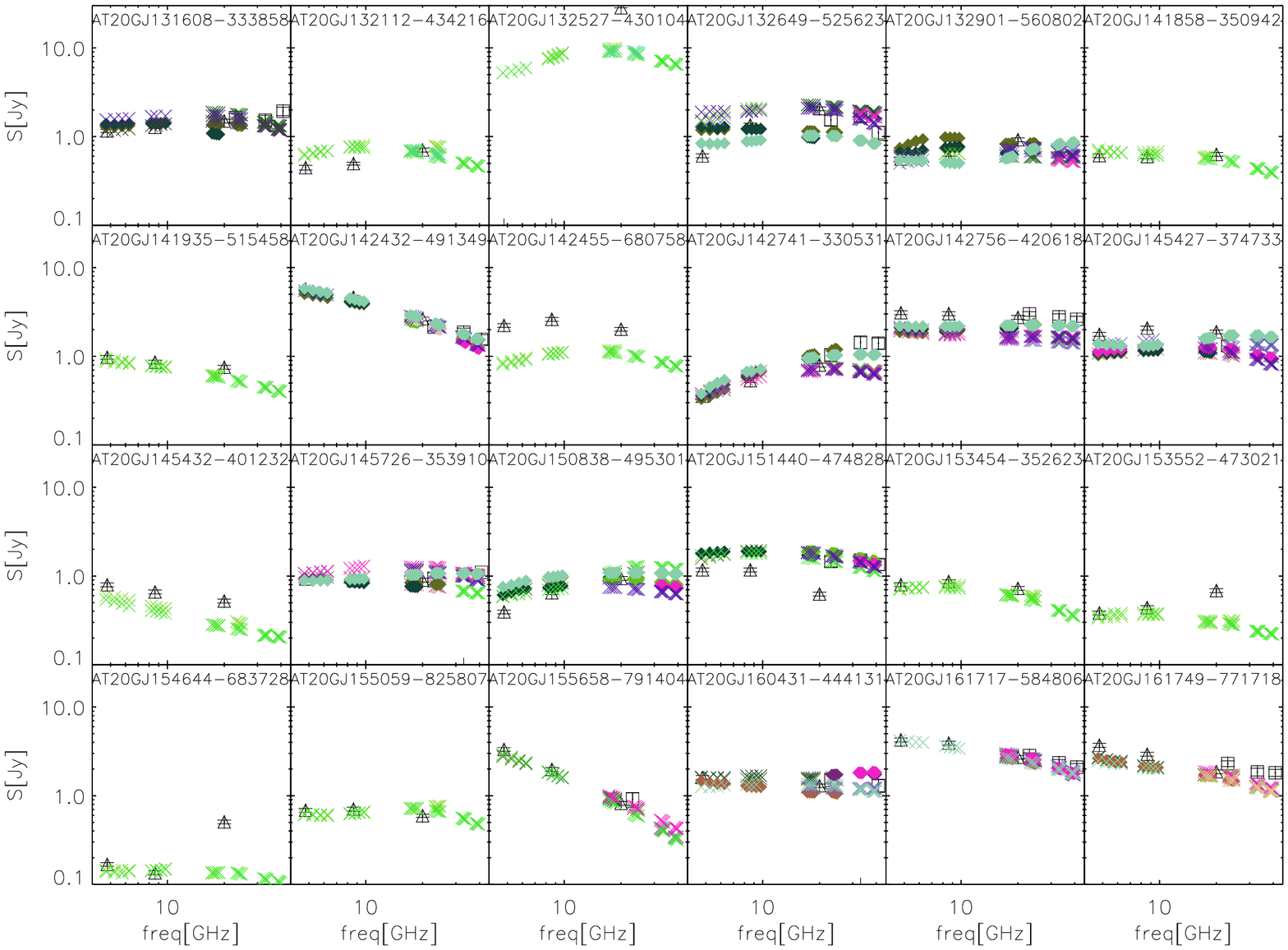}
  \includegraphics[width=11.5cm, angle=90]{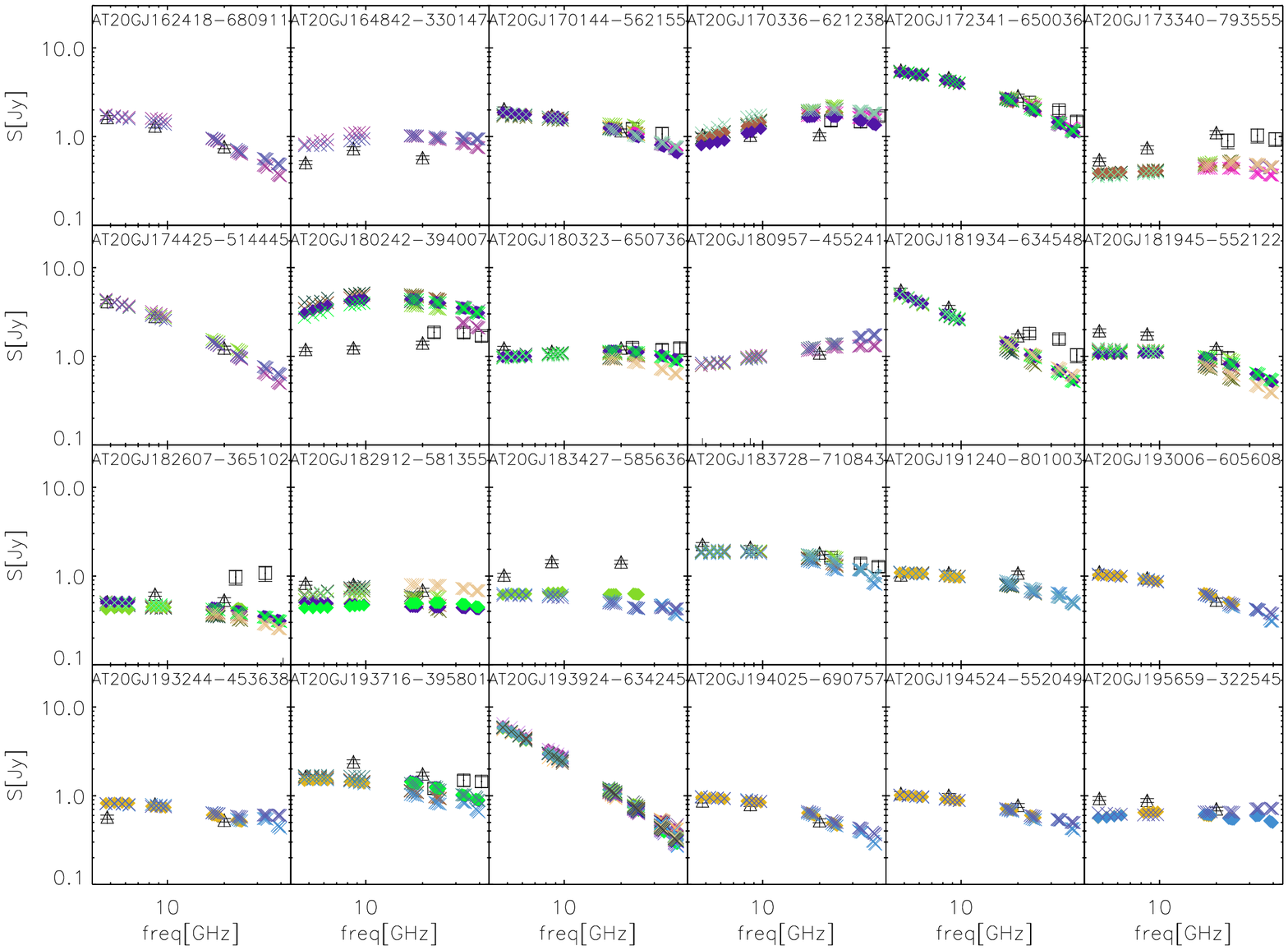}
  \caption{Continued.}
\end{figure*}
\addtocounter{figure}{-1}
\begin{figure*}
  \includegraphics[width=11.5cm, angle=90]{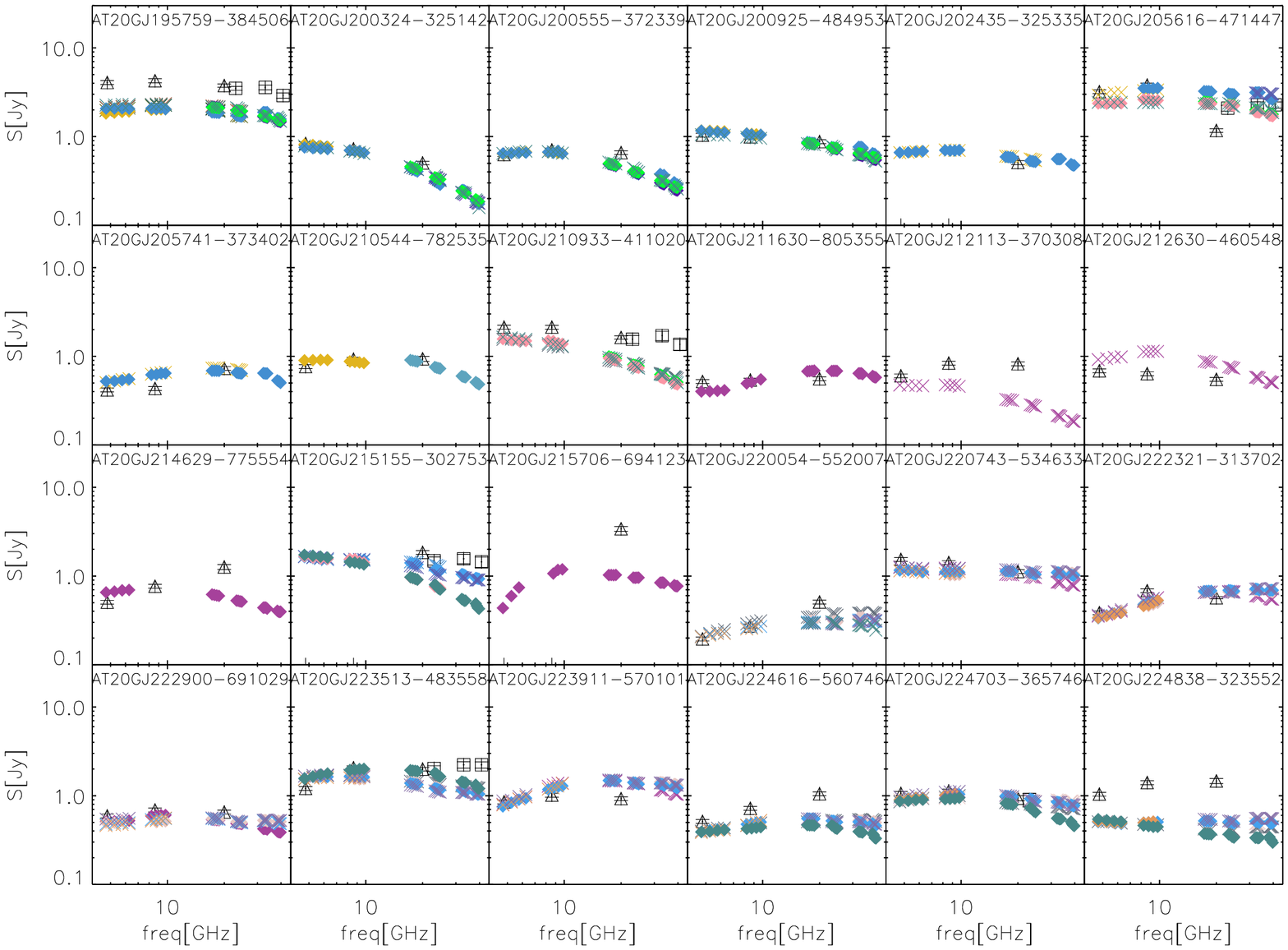}
  \includegraphics[width=7cm, angle=90]{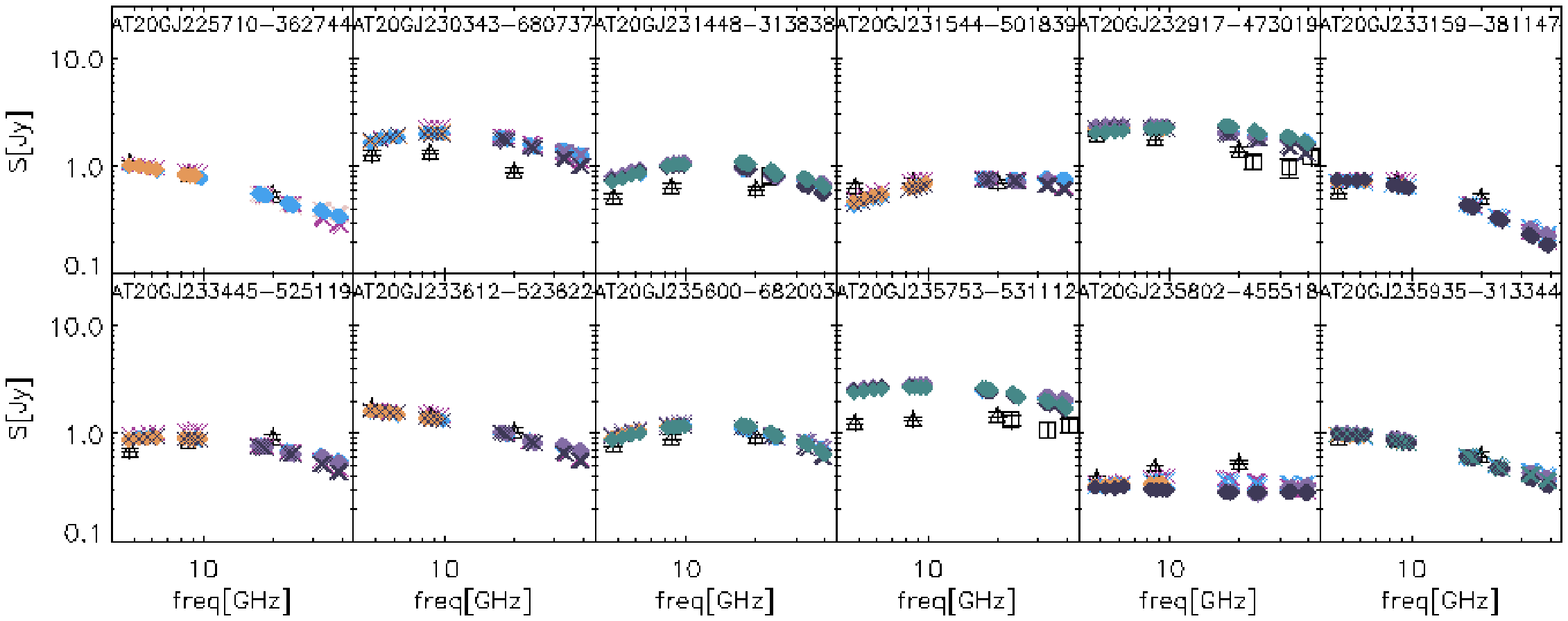}
  \caption{Continued.}
\end{figure*}

\section{The Bright Sources PACO catalogue} \label{sec:catalogue}

\begin{figure}
  \includegraphics[width=6cm]{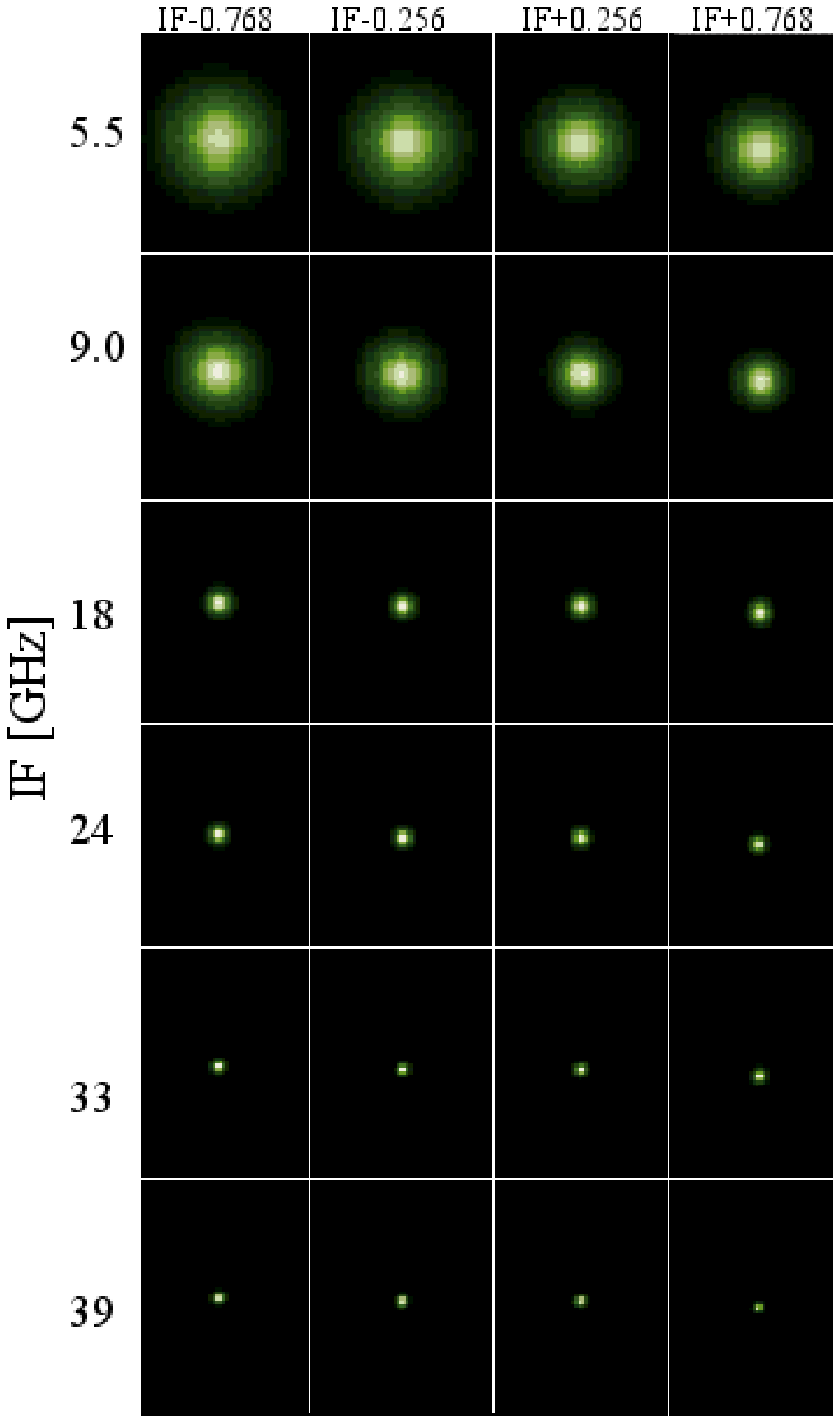}
  \caption{Images of the source AT20GJ053850-440508, observed on July 26th, 2009 with the hybrid configuration H75, as an example of the quality of our data and a representation of the beam. Each panel enclosing an image has a $\sim6'\times6'$ size. Rows show the images obtained in the four 512-MHz sub-bands associated to each of the 6 IFs.} \label{fig:image}
\end{figure}

Of the 189 sources comprised in the PACO bright sample, 13 are flagged as extended in the AT20G catalogue and 2 more (J062706-352916 and J130527-492804) were found to be extended by our 7mm data analysis. Two of the 13 AT20G extended sources (J051926-454554 and J051949-454643) correspond to the core and to the western lobe of Pictor~A. We have performed a dedicated observation in total intensity and polarization at 18, 24, 33 and 39 GHz of this well-known extended radio source by mosaicking the region including the core and the western lobe, which is the brightest and the more highly polarized (Burke-Spolaor et al. 2009). These observations will be presented in a separate  paper and the two sources have been left aside.

After the data reduction, 7 extended sources (J040848-750720, J084127-754028, J130527-492804, J161505-605427, J193557-462043, J195817-550923,J235904-605503) turned out to have unreliable flux density estimates at all the frequencies and for all the days of observations. This is mostly due to having used too extended array configurations, so that the sensitivity on the observed scale was poor. They have therefore not been included in the present catalogue.

For all the other extended objects (J052257-362730, J062706-352916, J074331-672625, J114531-483610, J132527-430104, J215706-694123) the estimated flux densities are lower limits to the integrated flux densities. Since these sources passed all the quality checks they have been included in the catalogue, with flux densities listed as lower limits, but have been excluded from further analysis.

Because of the large size of the table, the catalogue of the 180 sources for which we have a flux density estimation or a lower limit is available only online. Columns are as follows:
\begin{description}
\item[1] AT20G name;
\item[2-3] J2000 equatorial coordinates;
\item[4] epoch of observation as [yymmdd];
\item[5] flag `s' for ``simultaneity'' (i.e. within 10 days from the Planck observations). The epochs before February 2010 have to be compared with the first Planck survey, whereas the following epochs should be compared with the second Planck survey.
\item[6] flag `e' for extended sources;
\item[7-31] flux densities in mJy at the 24 frequencies between 4732 and 39768 MHz in which our 6$\times 2$-GHz bands have been divided;
\item[32-55] flux density errors;
\item[56-60] fit parameters, $S_0$, $\nu_0$, $a$, $b$ (see \S\,\ref{sec:spectra}).
\end{description}
Figure \ref{fig:SED} shows the spectra for the sources in the PACO bright sample. Different colours correspond to different observing epochs. Figure \ref{fig:image} shows, as an example, images at each frequency for the point source AT20G J053850-440508. For the epochs observed with hybrid arrays, as in this case, good quality images can be obtained even if we observe only one cut per source.
This is possible thanks to the brightness of the sources in our sample. It is a clear indication of the quality of our observations and of the beam size at the different frequencies.

\subsection{Spectral behaviour} \label{sec:spectra}

\begin{figure}
  \includegraphics[width=7cm, angle=90]{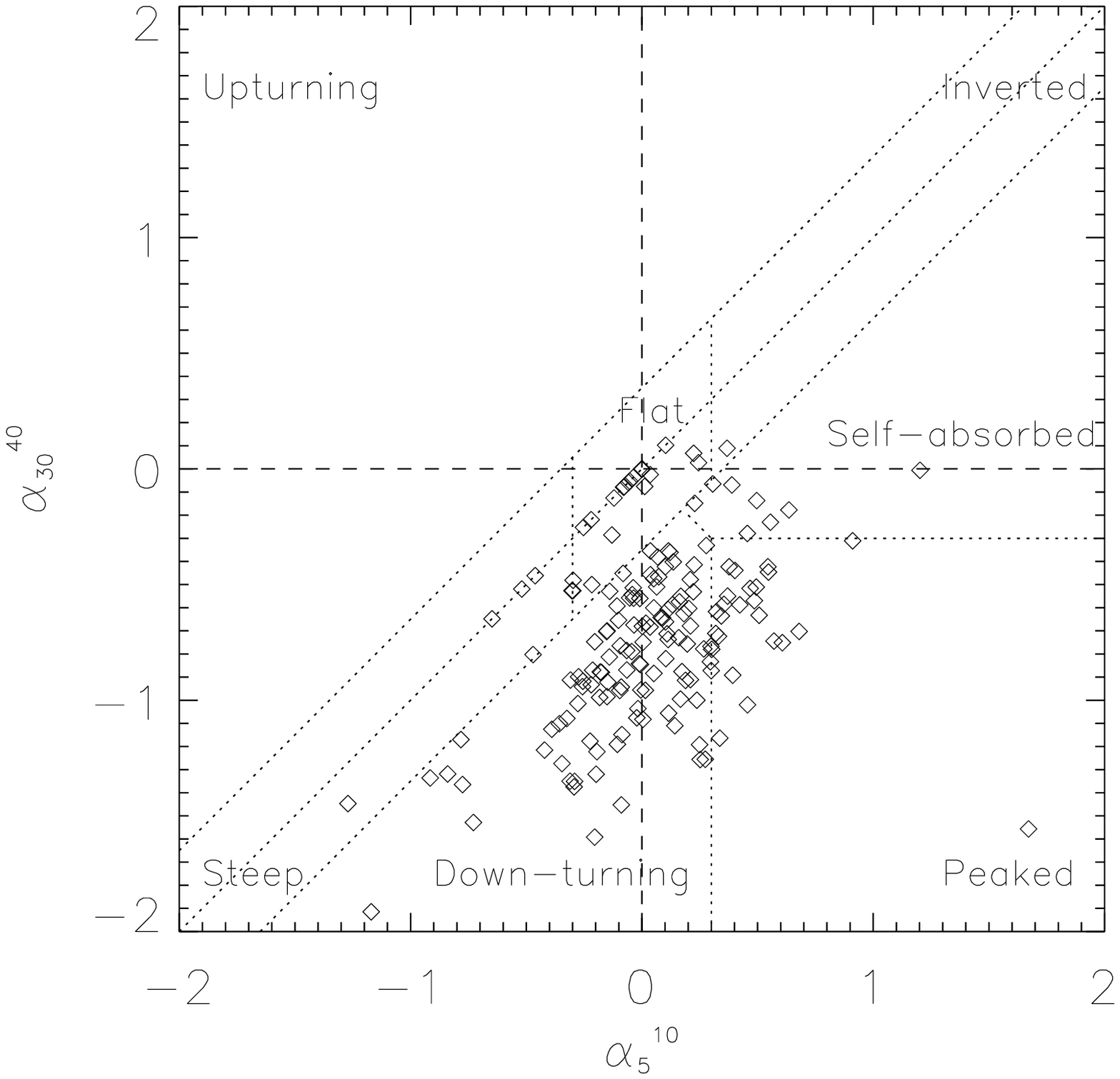}
  \caption{Colour-colour plot comparing the spectral indices in the ranges 5-10 and 30-40 GHz. The dotted lines shows the boundaries adopted for our spectral classification.} \label{fig:cc}
\end{figure}

Even though a power-law is a poor description for the majority of the spectra, conventional spectral indices are still useful for practical purposes. For example, they provide the easiest way to extrapolate the observed counts or model predictions from one frequency to another.

As our observations cover a wide frequency range (4.5-40 GHz), over which a single power-law is not enough to describe the spectral behaviour of the sources, we have studied the spectra of our 174 {\it point-like} sources by fitting the observed data with a double power law:
\begin{equation}
S(\nu)=S_0/[(\nu/\nu_0)^{-a}+(\nu/\nu_0)^{-b}]\label{dpl},
\end{equation}
where $\nu$ is the frequency, $S$ the flux density in Jy and $S_0$, $\nu_0$, $a$ and $b$ are free parameters. We considered only the sources for which we had at least 4 data points for each of the 2$\times 2\,$GHz bands considered. The fit has been performed in logarithmic units by minimizing the $\chi^2$  with a nonlinear optimization technique based on an implementation of the Generalised Reduced Gradient optimisation method (GRG, Windward Technologies, 1997). The allowed ranges for the parameters are: $4.5\,\hbox{GHz}\le\nu_0\le40\,\hbox{GHz}$, $a\ge -3 $, $b \le 3$. We have considered as bad fits those with $\chi^2-\langle \chi^2\rangle>3\sigma_\chi$ where $\langle \chi^2\rangle=1.3$ and $\sigma_\chi=0.54$ are mean and standard deviation of the Gaussian fit of the reduced $\chi^2$ distribution of the fits over the analyzed sample.

For 4 of the 174 sources we do not have enough data to obtain meaningful values of the parameters, while in 5 cases eq.~(\ref{dpl}) does not yield an acceptable fit ($\chi^2$ above the threshold). For the remaining 165 sources we have classified the spectral shape according to the values of the low- and high-frequency spectral indices $\alpha_{5}^{10}$ and $\alpha_{30}^{40}$ ($S\propto \nu^\alpha$) defined in the ranges 5-10 GHz and 30-40 GHz and calculated using the fitting formula. The distribution of these spectral indices is shown in Fig.~\ref{fig:cc}.

When $|\alpha_{5}^{10}-\alpha_{30}^{40}|<0.35$ the source spectrum is classified as `single power law'. We have 24 such sources ($\simeq 14\%$), further subdivided in:
 \begin{itemize}
 \item 6 (3.6\%) `steep' with $\alpha_{5}^{10}$, $\alpha_{30}^{40} <-0.3$,
 \item 17 (10.3\%) `flat' with $|\alpha_{5}^{10}|$, $|\alpha_{30}^{40}| \leq 0.3$,
 \item 1 (0.6\%) `inverted' with $\alpha_{5}^{10}$, $\alpha_{30}^{40} >0.3$.
 \end{itemize}
The other 141 have been classified as follows (see Fig.~\ref{fig:cc}):
 \begin{itemize}
 \item `peaked' if $\alpha_{5}^{10}>0.3$, $\alpha_{30}^{40} <-0.3$: 24 (14.5\%) sources;
 \item `down-turning' if $\alpha_{30}^{40}\le \min(\alpha_{5}^{10}-0.35, -\alpha_{5}^{10})$ and $\alpha_{5}^{10}\le 0.3$: 109 sources (66.0\%);
 \item  `self-absorbed' if $\alpha_{5}^{10}-0.35\ge \alpha_{30}^{40} \ge \max(-0.3, -\alpha_{5}^{10}$): 8 sources (4.8\%).
 \end{itemize}
Remarkably, we don't find any `upturning' source ($\alpha_{5}^{10}<\alpha_{30}^{40}$).
\begin{figure}
  \includegraphics[height=7cm, angle=90]{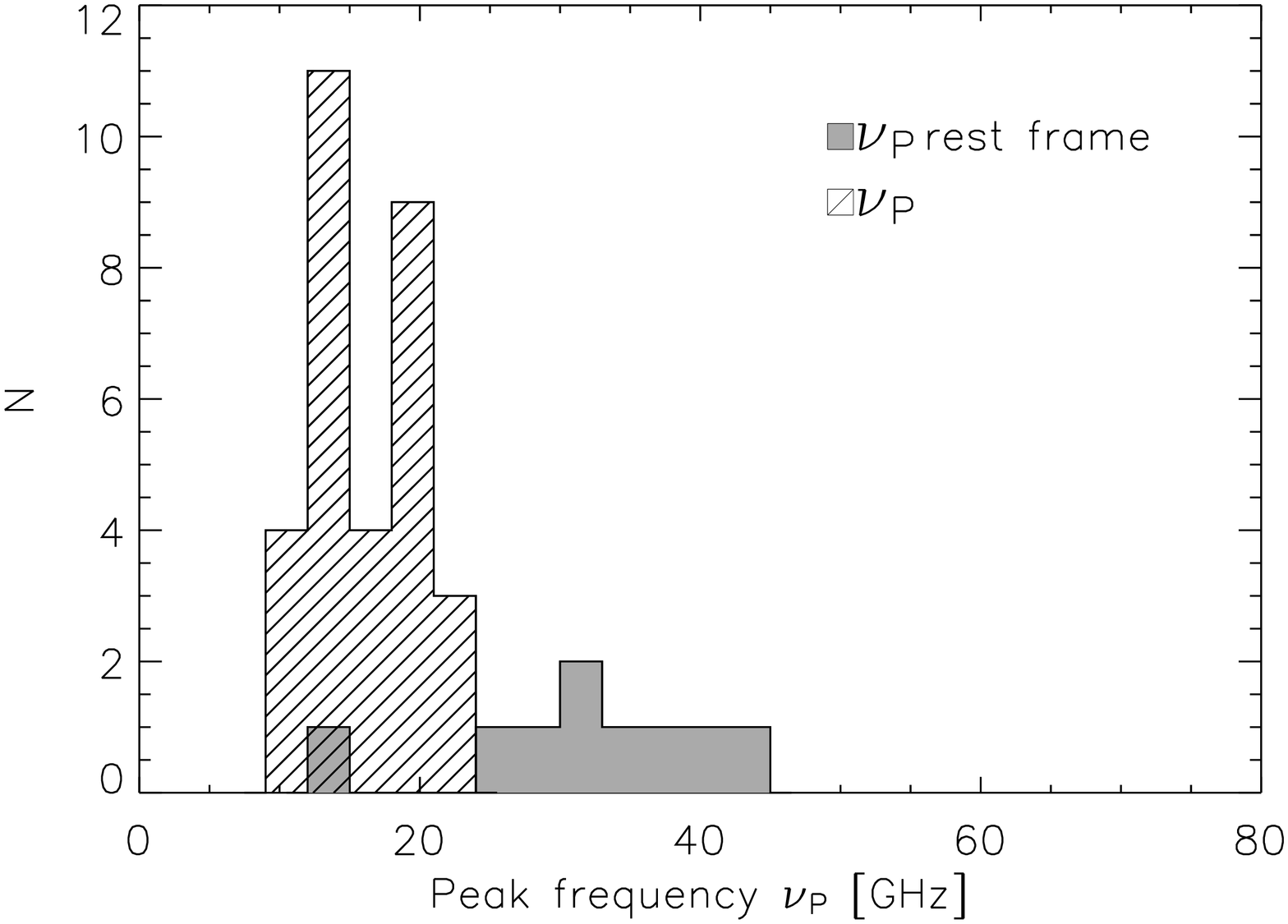}
  \caption{Distribution peak frequencies in the observer's frame (hatched histogram) and in the source frame (shaded) for sources with measured redshift. } \label{fig:peak_distr}
\end{figure}


\begin{figure}
  \includegraphics[height=7cm, angle=90]{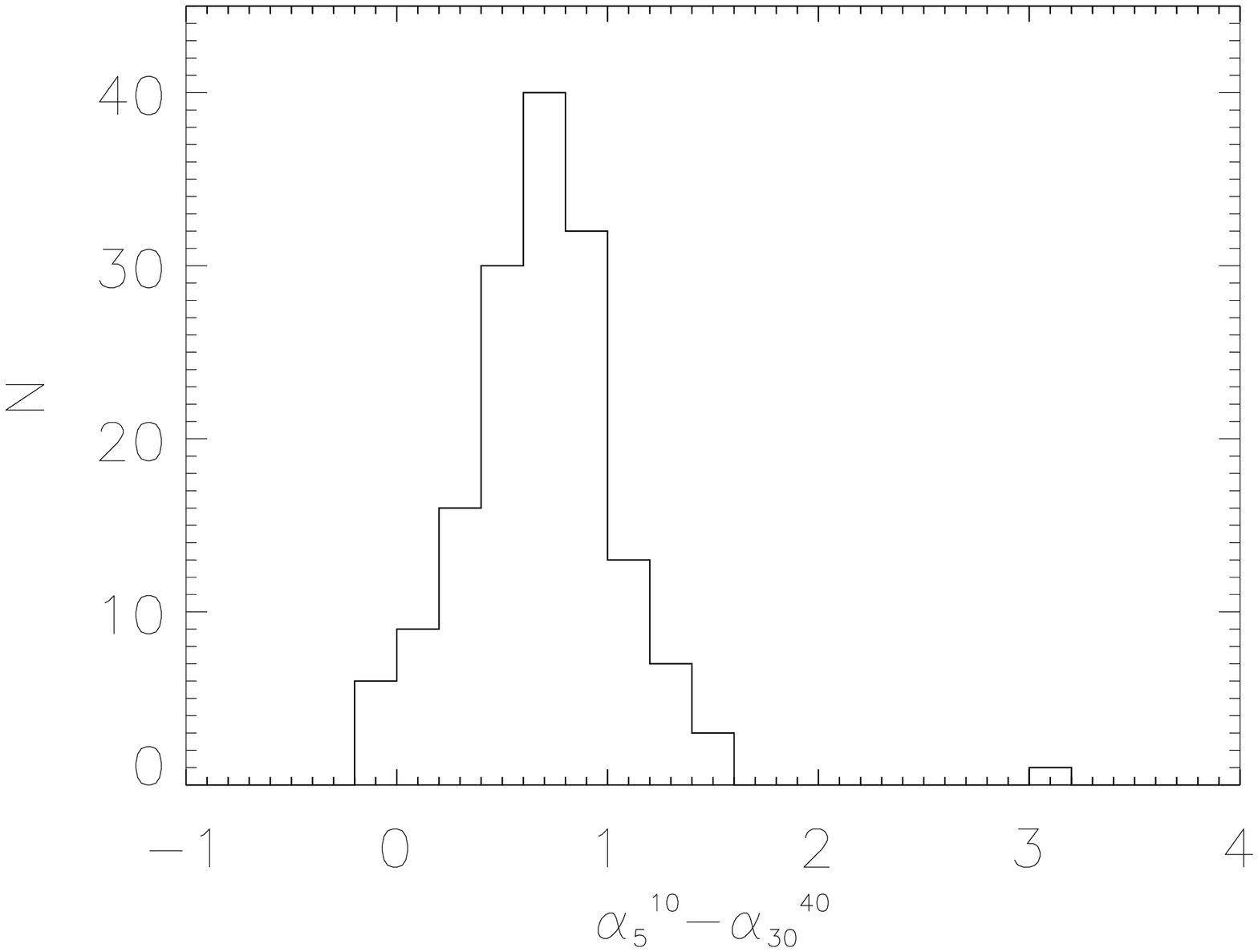}
  \caption{Distribution of the differences between low- and high-frequency spectral indices ($\alpha_{5}^{10}-\alpha_{30}^{40}$). } \label{fig:diff_sp_ind}
\end{figure}

\begin{figure}
  \includegraphics[height=7cm, angle=90]{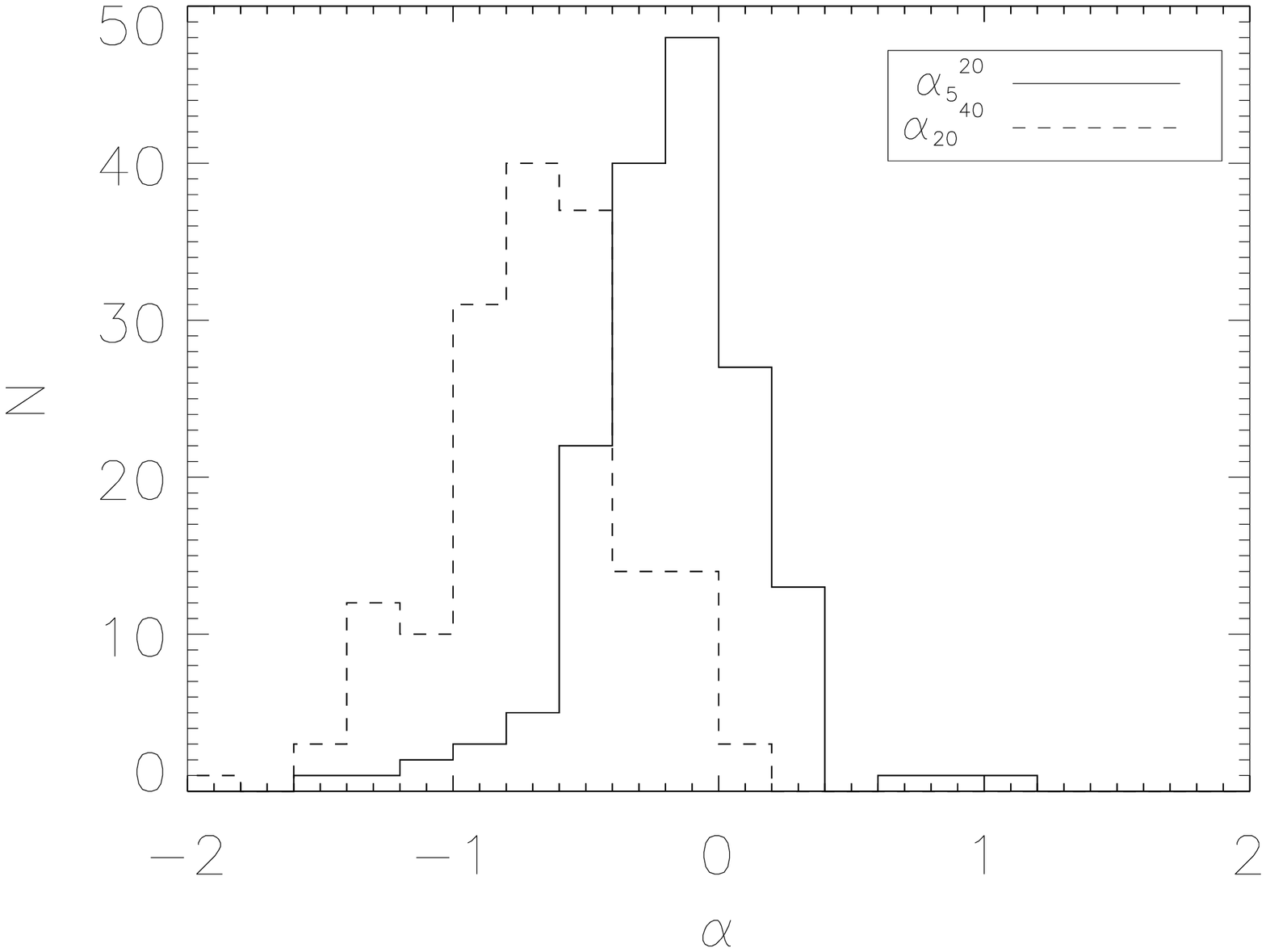}
  \caption{Distributions of spectral indices between 5 and 20 GHz (solid line) and between 20 and 40 GHz (dashed line). } \label{fig:sp_ind}
\end{figure}

The peak frequency of `peaked' sources is $\nu_p=\nu_0(-b/a)^{(1/(b-a))}$. The distribution of $\nu_p$ in the observer's frame is shown by the solid histogram in Fig.~\ref{fig:peak_distr}. The mean value is 16.4 GHz with a standard deviation of 3.5 GHz. The median is 15.5 GHz with inter-quartile range 14.0--19.7 GHz. The figure also shows the distribution of peak frequencies in the source frame for the 9 sources for which redshifts are available (Massardi et al. 2010, Mahony et al. in prep.). For these the mean is 32.1 GHz with dispersion of 8.8 GHz; the median is 32.2 GHz with inter-quartile range is 27.7--38.9 GHz.

The distribution of the differences between low- and high-frequency spectral indices ($\alpha_{5}^{10}-\alpha_{30}^{40}$) is shown in Fig.~\ref{fig:diff_sp_ind}. There is a clear steepening of the spectra at high frequency. The mean difference is 0.75 with a standard deviation of 0.37. The median difference is 0.73 with inter-quartile range 0.52--0.97.

Figure~\ref{fig:sp_ind} shows the distributions of spectral indices between 5 and 20 GHz and between 20 and 40 GHz (we refer to 20 GHz since it is the selection frequency of the PACO sample). The mean value of $\alpha_{5}^{20}$ is -0.07, with a standard deviation of 0.32, while the mean $\alpha_{20}^{40}$ is -0.55, with a standard deviation of 0.34. It should be noted that the 20 GHz selection favours sources that are brighter at this frequency. Therefore a positive value of $\alpha_{5}^{20}-\alpha_{20}^{40}$ is, at least partly, due to a selection effect.

Finally, Table~\ref{tab:spectindx} summarizes the median spectral indices calculated from the fitting double power-law function among couples of frequencies in the range covered by our observations. Since spectra are complex, spectral indices were computed only for relatively narrow frequency intervals.

\begin{table}
\caption{Matrix of median spectral indices calculated from the fitting double power-law function.}\label{tab:spectindx}
\begin{tabular}{c|ccccc}
\hline
Frequency [GHz]
     & 10   & 15    & 20    & 30 & 40  \\
\hline
5   & 0.04 & -0.02 & -0.07 & -  & -  \\
10  &  -    & -0.12 & -0.19 & -  & -  \\
15  &  -   &  -     & -0.25 & -0.38   & -  \\
20  &  -   &   -   &  -     & -0.45   & -0.55   \\
30  & -    &    -  &   -   &  -  & -0.70 \\
\hline
\end{tabular}
\end{table}

\subsection{Variability} \label{sec:variability}

\begin{table}
\begin{center}
\caption{Median variability index of the bright sample at different frequencies and time lags.}\label{tab:variability}
\begin{tabular}{lcccccc}
\hline
Time  & 5.5 & 9 & 18 & 24 & 33 & 39\\
range[d] & GHz & GHz & GHz & GHz & GHz & GHz \\
\hline
90 &   3.5&     6.3&     5.2&     5.7&     6.7&   7.5\\
180&   5.4&     5.8&     5.4&     7.6&     6.3&  7.5\\
270&   8.3&     5.0&     9.2&     8.4&    10.6&  11.3\\
\hline
\end{tabular}
\end{center}
\end{table}

Most of the sources in the sample have been observed more than once to monitor the flux density variations between the first two Planck surveys, carried out six months apart. Thus we typically re-observed each source after 6 months. More frequent monitoring was carried out only if spare time was available. Therefore observations with time lags shorter or longer than 6 months were done only for limited sub-samples.

Following Sadler et al. (2006) we define the variability index as:
$$
V_{\rm rms}=\frac{100}{\langle S \rangle}\sqrt{\frac{\sum [S_i-\langle S \rangle]^2-\sum\sigma^2_i}{n}},
$$
where $S_i$ is the flux density of a given source measured at the i-th epoch, $\sigma_i$ is the associated error, $n$ is the number of measurements, and $\langle S \rangle$ is the mean flux density of the source, computed using all the available observations of the source (not only those used for computing the variability index).

\begin{figure}
  \includegraphics[width=6cm, angle=90]{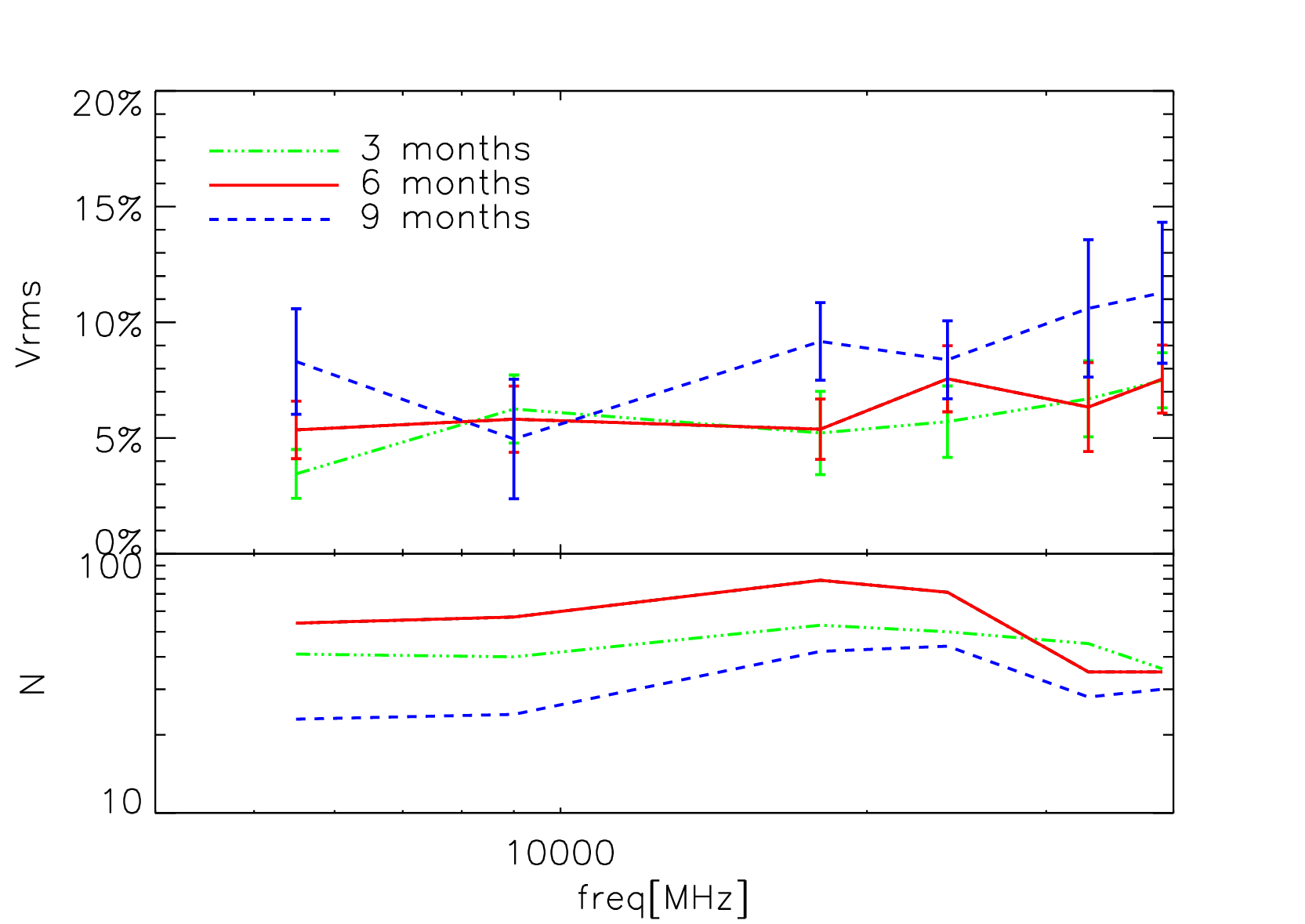}
  \caption{Top: median variability index of the sample as a function of frequency; error bars are $1.25\sigma/\sqrt{N-1}$. Bottom: number of sources.} \label{fig:var_index}
\end{figure}

To estimate the variability over a certain time interval we have selected for each source the best pair of observations (i.e., n = 2) spaced by the that time interval within $\pm 20\%$. In Fig~\ref{fig:var_index} we show the median variability index for the sample of sources as a function of frequency for time intervals of 3, 6, and 9 months (upper panel) and the number of sources having pairs of observations for those time intervals (lower panel).

We find, as expected (Impey \& Neugebauer 1988; Ciaramella et al. 2004), a trend toward an increase of the variability amplitude with frequency (see Table \ref{tab:variability} and Fig.~\ref{fig:var_index}). There is also a marginal indication of higher variability for the longer time lag, consistent with earlier results (e.g. Ciaramella et al. 2004). On the other hand, around 20 GHz the median variability index for a 9 months lag is $\simeq 9\pm 1.7\%$, slightly larger than that found by Sadler et al. (2006) over a one year timescale  (6.9\%) for a somewhat fainter sample (flux density limit of 100 mJy). The distribution of $V_{\rm rms}$, however, is similar to that found by Sadler et al.; for example, $37\pm 9\%$ of our sources show a variability index larger than 10\%, to be compared with $42\pm 7\%$ for the Sadler et al. sample.


\begin{table*}\tiny
\begin{center}
\caption{The 7\,mm flux densities, measured simultaneously with Planck observations, of  point-like calibrators in the PACO sample with  $S_{\rm AT20G}>5\,$Jy. The figure  immediately below each flux density measurement is the corresponding error.}\label{tab:calibs}
\begin{tabular}{llllcccccccc}
\hline
AT20G name & RA &$\delta$& Observing & $S_{\rm 32.2GHz}$&$S_{\rm 32.7GHz}$&$S_{\rm 33.2GHz}$&$S_{\rm 33.7GHz}$&$S_{\rm 38.2GHz}$&$S_{\rm 38.7GHz}$&$S_{\rm 39.2GHz}$&$S_{\rm 39.7GHz}$\\
   &$[hr]$&$[deg]$& Date &$[mJy]$&$[mJy]$&$[mJy]$&$[mJy]$&$[mJy]$&$[mJy]$&$[mJy]$&$[mJy]$\\
    &  && & $\sigma_{\rm S 32.2GHz}$&$\sigma_{\rm S 32.7GHz}$&$\sigma_{\rm S 33.2GHz}$&$\sigma_{\rm S 33.7GHz}$&$\sigma_{\rm S 38.2GHz}$&$\sigma_{\rm S 38.7GHz}$&$\sigma_{\rm S 39.2GHz}$&$\sigma_{\rm S 39.7GHz}$\\
   &&&  &$[mJy]$&$[mJy]$&$[mJy]$&$[mJy]$&$[mJy]$&$[mJy]$&$[mJy]$&$[mJy]$\\
\hline
   J053850-440508&   5.6473226& -44.0858154&  2010-02-28&  9506&  9483&  9385&  9356&  8962&  8921&  8864&  8893\\
&&&&   118&   118&   117&   117&   112&   111&   110&   111\\
   J053850-440508&   5.6473226& -44.0858154&  2010-03-15& 14278& 14210& 14135& 14137& 14848& 14620& 14639& 14814\\
&&&&   187&   186&   185&   185&   195&   192&   192&   194\\
   J053850-440508&   5.6473226& -44.0858154&  2009-09-15&  6344&  6352&  6339&  6339&  6192&  6156&  6177&  6141\\
&&&&    92&    92&    92&    92&    89&    89&    89&    89\\
   J125611-054721&  12.9364349&  -5.7893119&  2010-07-16& 14740& 14665& 14669& 14659& 14041& 13956& 14064& 14089\\
&&&&   126&   126&   126&   126&   120&   120&   121&   121\\
   J133739-125724&  13.6277171& -12.9568596&  2010-07-16&  4228&  4197&  4188&  4181&  3974&  3955&  3922&  3876\\
&&&&    36&    36&    36&    35&    34&    34&    33&    33\\
   J183339-210341&  18.5610881& -21.0611248&  2009-09-26&  3256&  3210&  3174&  3132&  2750&  2714&  2674&  2639\\
&&&&    16&    16&    16&    16&    14&    13&    13&    13\\
   J192451-291430&  19.4141825& -29.2417011&  2010-04-01& 17111& 17807& 17451& 17748&     0&     0&     0&     0\\
&&&&   196&   204&   200&   203&     0&     0&     0&     0\\
   J192451-291430&  19.4141825& -29.2417011&  2010-04-10& 14571& 14696& 14686& 14357& 14007& 13782& 13723& 13891\\
&&&&   213&   215&   215&   210&   205&   202&   201&   203\\
   J192451-291430&  19.4141825& -29.2417011&  2009-09-26& 14028& 13876& 13789& 13663& 12490& 12393& 12263& 12127\\
&&&&    71&    71&    70&    69&    63&    63&    62&    62\\
   J192451-291430&  19.4141825& -29.2417011&  2009-10-07& 15333& 15248& 15083& 14820& 14380& 14225& 14341& 14617\\
&&&&    60&    60&    59&    58&    56&    56&    56&    57\\
   J192451-291430&  19.4141825& -29.2417011&  2009-10-14& 13728& 13641& 13475& 13267& 12282& 12179& 12076& 12078\\
&&&&   122&   122&   120&   118&   110&   109&   108&   108\\
   J222547-045701&  22.4297933&  -4.9503870&  2009-11-06&  6177&  6098&  5929&  5905&  5557&  5337&  5009&  4978\\
&&&&   105&   103&   101&   100&    94&    90&    85&    84\\
   J222547-045701&  22.4297933&  -4.9503870&  2009-11-19&  5589&  5524&  5469&  5417&  4910&  4840&  4794&  4743\\
&&&&   110&   109&   108&   107&    97&    95&    94&    93\\
\hline
\end{tabular}
\end{center}
\end{table*}

\section{Absolute calibration for ground based telescopes} \label{sec:abscalib}

The absolute calibration of the Planck detectors up to 353 GHz is derived from the annual modulation of the CMB dipole by the satellite orbit around the Sun. An absolute measurement of the dipole is obtained by differentiating along a spin period. The CMB dipole direction is expected to be recovered to better than $\sim2$ arcmin, and its amplitude to better than $0.5\%$ (The Planck collaboration 2006). Planck aims at  obtaining an absolute photometric calibration better than 1\% for all frequency channels up to 353 GHz. As discussed in L\'opez-Caniego et al. (2007) the main limitation to the accuracy of flux density estimates of sources extracted from low-resolution maps (appropriate for CMB studies), such as these produced by WMAP and Planck, is confusion due to faint sources within the beam. Leach et al. (2008) estimated typical rms errors $\sim 100\,$mJy for the 30 and 44 GHz channels at high Galactic latitude. This corresponds to a few percent of the flux density of the brightest extragalactic sources.

Thus ground based observations, simultaneous with Planck's and at approximately the same frequencies, of very bright sources can be used to assess the absolute flux densities of calibrators for ground based telescopes. The PACO project has enough observations to determine the relative flux density scale between Planck and the ATCA.

Table~\ref{tab:calibs} lists the 7\,mm PACO flux densities of 6 compact sources with 20 GHz flux density $S_{\rm AT20G}>5\,$Jy and high quality observations simultaneous to Planck's. With the exception of AT20GJ053850-440508, these sources lie outside the region covered by the bright sample. All of them are well-known highly variable blazars. Two of them are equatorial sources and are currently used in some ad-hoc observational runs for cross-calibration between southern and northern hemisphere telescopes, in order to assure that the Planck absolute calibration could be shared with all the major ground-based facilities.

As noted before, the ERCSC (Planck collaboration 2011b) fluxes are unsuitable for calibrating ground-based telescopes. A proper comparison with Planck flux density scales will be possible, and will be done, when Planck time ordered data will become available. A preliminary comparison between ground based and Planck measurements has been presented by Planck collaboration (2011b, 2011e).

\section{Conclusions} \label{sec:conclusion}

The PACO project consists in observations with ATCA of several complete samples in three pairs of 2 GHz bands (5.5 and 9 GHz, 18 and 24 GHz, 33 and 39 GHz). The observations were carried out in several epochs between July 2009 and August 2010. At least one observation was made within 10 days from the Planck satellite observations in any of the LFI channels (30, 44, 70 GHz).

We have presented and analyzed the data on sources in the ``bright sample'' ($S_{20\rm GHz}>500\,$mJy). Although the present analysis is preliminary, in wait for the much broader spectral coverage that will be possible combining our measurements with Planck data, we note some interesting results:
\begin{itemize}
\item Only a minor fraction ($\simeq 14\%$) of source spectra have a shape close to a single power-law over the 4.5 to 40 GHz range; they are mostly ``flat''-spectrum.
\item Few sources have a positive (i.e. rising, $S\propto \nu^\alpha$) high-frequency (30 to 40 GHz) spectral index, and none has an upturning spectrum ($\alpha_{5\rm GHz}^{10\rm GHz}<\alpha_{30\rm GHz}^{40\rm GHz}$). Most sources show a spectral steepening above 10--20 GHz: the mean value of the spectral index is $\simeq -0.07$ (with a standard deviation of 0.35) between 5 and 20 GHz and decreases to $\simeq -0.57$ (with a standard deviation of 0.32) between 20 and 40 GHz. Break frequencies are almost uniformly distributed over the full frequency range covered by our observations. We do not find any correlation between the variation of the spectral index and the break frequency. We note that part of this steepening may be due to a selection effect since the primary 20 GHz selection of the sample emphasizes sources which are brighter at this frequency.
\item A significant fraction (about 14\%) of the sources show a peak within our frequency range. The mean value of the peak frequency is 16.3 GHz with a standard deviation of 3.5 GHz. The mean rest frame peak frequency for the 9 such sources for which a redshift is available is 32.1 GHz with dispersion of 8.8 GHz. In general, however, the peak is not very prominent.
\item A handful of sources (8, i.e. 4.7\%) show a rising low-frequency spectrum, followed by a flattening, perhaps indicative of self-absorption up to $\simeq 10\,$GHz.
\item There is a trend toward an increase of the variability amplitude with frequency and a marginal indication of higher variability for the longer time lag.
\end{itemize}

We also present accurate ATCA flux density measurements, simultaneous with Planck, for 6 bright ($S_{\rm AT20G}>5\,$Jy) point-like sources that can be used to exploit the excellent absolute calibration achieved by Planck to re-assess the high frequency calibration of ground-based telescopes and to improve models for planets and other sources used as calibrators at mm wavelengths, once the satellite time ordered data are available.

On the other hand, our data will help in the validation of sources detected by Planck and in quantifying the effect of confusion on Planck flux density and position estimates.

\section*{Acknowledgments}
MM, AB, and GDZ acknowledge financial support from ASI (ASI/INAF Agreement I/072/09/0 for the Planck LFI activity of Phase E2 and contract I/016/07/0 'COFIS'). MLC acknowledges partial financial support from the Spanish Ministerio
de Ciencia e Innovaci\'on proyects AYA2010-21766-C03-01 and CSD2010-00064.
RDE acknowledges support from an Australian Federation Fellowship (FF0345330).

We thank the staff at the Australia Telescope Compact Array site, Narrabri (NSW), for the valuable support they provide in running the telescope. In particular we are indebted to Philip G. Edwards for the unvaluable help in scheduling the ATCA runs according to the Planck scanning strategy and to James Stevens for the suggestions about the data reduction pipeline. The Australia Telescope Compact Array is part of the Australia Telescope which is funded by the Commonwealth of Australia for operation as a National Facility managed by CSIRO.

\bsp

\label{lastpage}

\end{document}